\documentclass[aps,prd,twocolumn,showpacs,floatfix,byrevtex]{revtex4-1}
\usepackage{graphicx}
\usepackage{dcolumn}
\usepackage{amsmath}
\usepackage{epsfig}
\usepackage{color}

\newcommand{\be}{\begin{enumerate}}
\newcommand{\ee}{\end{enumerate}}

\newcommand{\etal}{\it et al.\rm}

\newcommand\T{\rule{0pt}{2.6ex}}       
\newcommand\B{\rule[-1.2ex]{0pt}{0pt}} 

\begin{document}
 
\title{\boldmath Branching fraction measurements of $\psi(3686) \to
  \gamma \chi_{cJ}$}

\author{
\small
M.~Ablikim$^{1}$, M.~N.~Achasov$^{9,e}$, S. ~Ahmed$^{14}$, M.~Albrecht$^{4}$, A.~Amoroso$^{50A,50C}$, F.~F.~An$^{1}$, Q.~An$^{47,a}$, J.~Z.~Bai$^{1}$, O.~Bakina$^{24}$, R.~Baldini Ferroli$^{20A}$
% ...
\\
\vspace{0.2cm}
(BESIII Collaboration)\\
\vspace{0.2cm} {\it
$^{1}$ Institute of High Energy Physics, Beijing 100049, People's Republic of China\\
% ...
$^{4}$ Bochum Ruhr-University, D-44780 Bochum, Germany\\
% ...
$^{9}$ G.I. Budker Institute of Nuclear Physics SB RAS (BINP), Novosibirsk 630090, Russia\\
% ...
$^{14}$ Helmholtz Institute Mainz, Johann-Joachim-Becher-Weg 45, D-55099 Mainz, Germany\\
% ...
$^{20}$ (A)INFN Laboratori Nazionali di Frascati, I-00044, Frascati, Italy; (B)INFN and University of Perugia, I-06100, Perugia, Italy\\
% ...
$^{24}$ Joint Institute for Nuclear Research, 141980 Dubna, Moscow region, Russia\\
% ...
$^{47}$ University of Science and Technology of China, Hefei 230026, People's Republic of China\\
% ...
$^{50}$ (A)University of Turin, I-10125, Turin, Italy; (B)University of Eastern Piedmont, I-15121, Alessandria, Italy; (C)INFN, I-10125, Turin, Italy\\
% ...
$^{a}$ Also at State Key Laboratory of Particle Detection and Electronics, Beijing 100049, Hefei 230026, People's Republic of China\\
% ...
$^{e}$ Also at the Novosibirsk State University, Novosibirsk, 630090, Russia\\
% ...
}
\vspace{0.4cm}
}




\begin {abstract}
  Using a sample of 106 million $\psi(3686)$ decays, the branching
  fractions of $\psi(3686) \to \gamma \chi_{c0}, \psi(3686) \to \gamma
  \chi_{c1}$, and $\psi(3686) \to \gamma \chi_{c2}$ are measured with
  improved precision to be $(9.389 \pm 0.014 \pm 0.332)\,\%$, $(9.905 \pm
  0.011 \pm 0.353)\,\% $, and $(9.621 \pm 0.013 \pm 0.272)\,\% $,
  respectively, where the first uncertainties are statistical and the
  second ones are systematic.  The product branching fractions of
  $\psi(3686) \to \gamma \chi_{c1}, \chi_{c1} \to \gamma J/\psi$ and
  $\psi(3686) \to \gamma \chi_{c2}, \chi_{c2} \to \gamma J/\psi$ and
  the branching fractions of $\chi_{c1} \to \gamma J/\psi$ and
  $\chi_{c2} \to \gamma J/\psi$ are also presented.
\end{abstract}

\date{\today}

\maketitle


\section{Introduction}

The discovery of the $J/\psi$ in 1974 and soon thereafter of
the charmonium family convinced physicists of the reality of the quark
model~\cite{quarks}.  Since then, measurements of the masses
and widths of the charmonium family and their hadronic and radiative
transition branching fractions have become more precise.  The spectrum
of bound charmonium states is important for the understanding of
Quantum Chromodynamics (QCD) in the perturbative and non-perturbative
regions~\cite{M1}.

For charmonium states that are above the ground state but below
threshold for strong decay into heavy flavored mesons, like the
$\psi(3686)$, electromagnetic decays are important decay modes.  The
first charmonium states discovered after the $J/\psi$ and $\psi(3686)$
were the $\chi_{cJ}$ ($J$ = 0, 1, and 2) states, which were found in
radiative transitions of the $\psi(3686)$~\cite{disc1,disc2}.  These
states, which are the triplet $1P$ states of the $c\bar{c}$ system,
had been theoretically predicted~\cite{187,188} along with the
suggestion that they could be produced by $E1$ transitions from the
$\psi(3686)$ resonance.

Radiative transitions are sensitive to the inner structure of
hadrons, and experimental progress and theoretical progress are important for
understanding this structure.  The development of theoretical models
is also important for predicting the properties of missing charmonium
states, in order to help untangle charmonium states above the open-charm
threshold from the mysterious $XYZ$ states~\cite{PDG16}.  Much information on
radiative transitions of charmonium can be found in Ref.~\cite{M1},
and a recent summary of theoretical predictions for radiative
transitions of charmonium states and comparisons with experiment may
be found in Ref.~\cite{Deng1}.

The branching fractions of $\psi(3686) \to \gamma \chi_{cJ}$ were
measured most recently by CLEO in 2004 with a sample of 1.6 M
$\psi(3686)$ decays~\cite{cleo04}.  The Crystal Ball~\cite{gaiser}, CLEO
values, and the Particle Data Group (PDG)~\cite{PDG16} averages are
given in Table~\ref{CLEO_results}.

\begin{table}[h]
\centering
\begin{footnotesize}
  \caption{Crystal Ball~\cite{gaiser} and CLEO~\cite{cleo04}
    $\psi(3686) \to \gamma \chi_{cJ}$ branching fractions and
    average values from the PDG~\cite{PDG16}.}
\vspace{0.05in}
\begin{tabular}{l|c|c|c} \hline \hline
\T Decay & Crystal Ball (\%) & CLEO (\%) & PDG (\%) \B \\ \hline
\T $\psi(3686) \to \gamma \chi_{c0}$ &  $9.9 \pm 0.5
\pm 0.8$ & $9.22 \pm 0.11 \pm 0.46$ &   $9.2 \pm 0.4$  \\  
$\psi(3686) \to \gamma \chi_{c1}$ & $9.0 \pm 0.5
\pm 0.7$ &  $9.07 \pm 0.11 \pm 0.54$ & $8.9 \pm 0.5$\\  
$\psi(3686) \to \gamma \chi_{c2}$ & $8.0 \pm 0.5
\pm 0.7$ & $9.33 \pm 0.14 \pm 0.61$ &  $8.8 \pm 0.5$ \B \\  
\hline \hline
\end{tabular}
\end{footnotesize}
\label{CLEO_results}
\end{table}

BESIII has the world's largest sample of $\psi(3686)$ decays and has
made precision measurements of many $\psi(3686)$ branching fractions,
including $\psi(3686) \to \pi^+ \pi^- J/\psi$, along with $J/\psi \to
l^+ l^-$ ($l = e, \mu$)~\cite{pipij}, $\psi(3686) \to \pi^0 J/\psi$
and $\eta J/\psi$~\cite{pi0jpsi}, $\psi(3686) \to \pi^0
h_c$~\cite{brhc1,brhc2}, and the product branching fractions
$\mathcal{B}(\psi(3686) \to \gamma \chi_{cJ})\times
\mathcal{B}(\chi_{cJ} \to \gamma J/\psi)$~\cite{xiaorui,bam158} using
exclusive $\chi_{cJ} \to \gamma J/\psi$ decays.  It is important that
the $\psi(3686) \to \gamma \chi_{cJ}$ and $\psi(3686) \to \gamma
\eta_c$ branching fractions be measured as well. Improved precision on
these is necessary because they are often used in the determination of
$\chi_{cJ}$ and $\eta_c$ branching fractions via the product branching
fractions.  However, it is to be noted that systematic uncertainties
dominate the measurements summarized in Table~\ref{CLEO_results}, so
to improve on their results, it is necessary to reduce the systematic
uncertainties.

In this paper, we analyze $\psi(3686)$ inclusive radiative decays and
report the measurement of the $\psi(3686) \to \gamma \chi_{cJ}$
branching fractions.  The product branching fractions
$\mathcal{B}(\psi(3686) \to \gamma \chi_{cJ})\times
\mathcal{B}(\chi_{cJ} \to \gamma J/\psi)$ are also measured, and the
$\chi_{cJ} \to \gamma J/\psi$ branching fractions are determined.
This analysis is based on the $\psi(3686)$ event sample taken in 2009 of
106 million events, determined from the number of hadronic decays as
described in Ref. [17], the corresponding continuum sample with
integrated luminosity of 44 pb$^{-1}$ at $\sqrt{s} =
3.65$~GeV~\cite{Npsip}, and a 106 million $\psi(3686)$ inclusive Monte
Carlo (MC) sample.

The paper is organized as follows: In Section II, the BESIII detector
and inclusive $\psi(3686)$ MC simulation are described. In Section
III, the selections of inclusive $\psi(3686) \to \gamma X$ events and
$\pi^0$'s are described and comparisons of inclusive $\psi(3686)$ data
and MC sample distributions are made. Section IV presents the
inclusive photon energy distributions, while Section V details the
selection of exclusive $\psi(3686) \to \gamma \chi_{cJ}$ events.
Sections VI and VII describe the fitting of the photon energy
distributions and the determination of the branching fractions,
respectively. Section VIII presents the systematic uncertainties, and
Sections IX and X give the results and summary, respectively.

\section{\boldmath BESIII and Inclusive $\psi(3686)$ Monte Carlo Simulation}
BESIII is a general-purpose detector at the double-ring
$e^+e^-$ collider BEPCII and is used for the study of physics in
the $\tau$-charm energy region~\cite{taucharm}. It has a geometrical
acceptance of 93\,\% of $4\pi$ solid angle and consists of four main
subsystems: a helium-based multi-layer drift chamber (MDC), a plastic
scintillator time-of-flight~(TOF) system, a CsI(Tl) electromagnetic
calorimeter~(EMC) and a resistive plate muon chamber system. The first
three sub-detectors are enclosed in a superconducting solenoidal
magnet with a 1.0 T magnetic field. More details of the detector are
described in Ref.~\cite{Ablikim2010345}.

MC simulations of the full detector are used to determine detection
efficiency and to understand potential backgrounds. The {\sc
  geant4}-based~\cite{GEANT4} simulation software, BESIII Object
Oriented Simulation~\cite{BOOST}, contains the detector geometry and
material description, the detector response and signal digitization
models, as well as records of the detector running conditions and
performance. Effects of initial state radiation~(ISR) are taken into
account with the MC event generator {\sc kkmc}~\cite{kkmc1,kkmc2}, and
final state radiation~(FSR) effects are included in the simulation by
using PHOTOS~\cite{PHOTOS}.  Particle decays are simulated with {\sc
  evtgen}~\cite{evtgen} for the known decay modes with branching
fractions set to the world average~\cite{PDG16} and with the {\sc
  lundcharm} model~\cite{lund} for the remaining unknown decays.

Angular distributions of the cascade $E1$ transitions $\psi(3686) \to
\gamma \chi_{cJ}$ follow the formulas in Refs.~\cite{Karl,liao}, while
the $\cos \theta$ distributions for $\chi_{cJ} \to \gamma J/\psi$ are
generated according to phase space distributions.  The $\chi_{cJ}$ are
simulated with Breit-Wigner line shapes.  To account for the $E1$
transitions for $\psi(3686) \to \gamma \chi_{cJ}, \chi_{cJ} \to \gamma
J/\psi$, MC events will be weighted as described in
Section~\ref{inclusive-section}.

\section{\boldmath Event Selection}
\subsection{\boldmath Inclusive $\psi(3686) \to \gamma X$ Event
  Selection}
\label{inc_sel}
We start by describing the selection procedure for $\psi(3686)$ event
candidates.  To minimize systematic uncertainties from selection
requirements, the $\psi(3686)$ event selection criteria, which are
used for both data and the MC sample, are fairly loose.

Charged tracks must be in the active region of the MDC with $|\cos
\theta| < 0.93$, where $\theta$ is the polar angle of the track, and
have $V_r < 2$ cm and $|V_z| < 10$ cm, where $V_r$ is the distance of
the point of closest approach of the track to the beam line in the
plane perpendicular to the beam line and $|V_z|$ is the distance to
the point of closest approach from the interaction point along the
beam direction. In addition, $p < 2.0$ GeV$/c$ is required to eliminate
misreconstructed tracks, where $p$ is the track momentum.

Photon candidates are reconstructed from clusters of energy in the EMC
that are separated from the extrapolated positions of any charged
tracks by more than 10 standard deviations and have reconstructed
energy $E_{\gamma} > 25$ MeV in the EMC barrel ($|\cos
\theta_{\gamma}| < 0.80$) or $ > 50$ MeV in the EMC end-caps ($0.86 <
|\cos \theta_{\gamma}| < 0.92$), where $E_{\gamma}$ is the photon
energy and $\theta_{\gamma}$ is the polar angle of the photon.  The
energy deposited in nearby TOF counters is included in EMC
measurements to improve the reconstruction efficiency and energy
resolution.  Photons in the region between the barrel and end-caps are
poorly reconstructed and are not used.  In addition, $E_{\gamma} <
2.0$ GeV is required to eliminate misreconstructed photons.  The
timing of the shower is required to be no later than 700 ns after the
reconstructed event start time to suppress electronic noise and energy
deposits unrelated to the event.

To help in the selection of good $\psi(3686)$ candidates, events must
have $N_{\text{ch}} > 0$, where $N_{\text{ch}}$ is the number of
charged tracks, and $E_{\text{vis}} = E_{\text{ch}} + E_{\text{neu}} >
0.22E_{\text{cm}}$, where $E_{\text{vis}}$ is the visible energy of
the event, $E_{\text{ch}}$ is the total energy of the charged
particles assuming them to be pions, $E_{\text{neu}}$ is the total
energy of the photons in the event, and $E_{\text{cm}}$ is the center
of mass (CM) energy. To remove beam background related showers in the
EMC and to demand at least one photon candidate in order to select
inclusive $\psi(3686) \to \gamma X$ events, we require $0 < N_{\gamma}
< 17$, where $N_{\gamma}$ is the number of photons.  In the following,
inclusive $\psi(3686)$ events and inclusive $\psi(3686)$ MC events
will assume this selection.

\subsection{\boldmath Non-$\psi(3686)$ background}
\label{non_psi_bkg}
By examining the continuum sample taken at a CM energy of
3.65 GeV, a set of selection requirements were chosen to further
remove non-$\psi(3686)$ background by identifying Bhabha events,
two-photon events, ISR events, beam background events, electronic
noise, etc.  Events satisfying any of the following conditions will
be removed:
\begin{enumerate}
\item $N_{\text{ch}} <4$ and $p_{i}c > 0.92 E_{\text{beam}}$, where $E_{\text{beam}}$ is the beam energy and the
$p_i$ is the momentum of any charged track in the event.
\item  $N_{\text{ch}} <4$ and $(E_{\text{EMC}})_{i} > 0.9 E_{\text{beam}}$, where  $(E_{\text{EMC}})_i$ is the
deposited energy of any charged or neutral track in the EMC.
\item $N_{\text{ch}} <4$ and $E_{\text{cal}} < 0.15 E_{\text{cm}}$,
  where $E_{\text{cal}}$ is the total deposited energy (charged
  and neutral) in the $EMC$.
\item $N_{\text{ch}} = 1$ and $(E_{\text{ch}} + E_{\text{neu}}) < 0.35 E_{\text{cm}}$. 
\item $|((P_z)_{\text{ch}} + (P_z)_{\text{neu}})|c > 0.743E_{\text{beam}}$, where ($P_z)_{\text{ch}}$
and $(P_z)_{\text{neu}}$ are the sums of the momenta of the charged and
neutral tracks in the $z$ direction.
\end{enumerate}
CLEO in Ref.~\cite{cleo04} used a similar selection in their
analysis.

\subsection{\boldmath $\pi^0$ candidate selection}
\label{good-pi0}
The invariant mass distribution of all $\gamma \gamma$ combinations
has a clear peak from $\pi^0 \to \gamma \gamma$ decay.  To reduce
background under the radiative transition peaks, photons in $\pi^0$'s
will be removed from the inclusive photon energy distributions.  To
reduce the loss of good radiative transition photons due to accidental
mis-combinations under the $\pi^0$ peak, the requirements for a
$\pi^0$ candidate are rather strict.

Photons in $\pi^0$ candidates must have $\delta> 14$ degrees, where
$\delta$ is the angle between the photon and the closest charged track
in the event, and the lateral shower profile must be consistent with
that of a single photon.  The $\pi^0$ candidates must have at least
one photon in the EMC barrel; a one-constraint kinematic fit to the
nominal $\pi^0$ mass with a  $\chi^2 < 200$; and $ 0.12 <
M_{\gamma \gamma} < $ 0.145 GeV$/c^2$, where $M_{\gamma \gamma}$ is
the $\gamma \gamma$ invariant mass.  In addition, $|\cos \theta^*| <
0.84$ is required for a $\pi^0$ candidate, where $\theta^*$ is the
angle of a photon in the $\pi^0$ rest frame with respect to the
$\pi^0$ line of flight.  Real $\pi^0$ mesons decay isotropically, and
their decay angular distribution is flat. However $\pi^0$ candidates
that originate from a wrong photon combination do not have a flat
distribution and peak near $|\cos \theta^*| = 1$.

\subsection{\boldmath Comparison of inclusive $\psi(3686)$ data and
  the MC sample}
\label{compare}
Since efficiencies and backgrounds depend on the accuracy of the MC
simulation, it is important to validate the simulation by comparing
the inclusive $\psi(3686)$ MC with on-peak data minus continuum data.
In the following, data will refer to on-peak data minus scaled
continuum data, where the scale factor of 3.677 accounts for the
difference in energy and luminosity between the two data
sets~\cite{Npsip}.  In general, data distributions compare well with
the inclusive MC distributions, except for those involving $\pi^0$s.
To improve the agreement, each MC event is given a weight determined
by the number of $\pi^0$s, $N_{\pi^0}$, in the event.  For events with
$N_{\pi^0}$ corresponding to bin $i$ of the $N_{\pi^0}$ distribution,
$w_{\pi^0} = \frac{(N_{\pi^0}^{\text{data}})_i}{(N_{\pi^0}^{\text{MC}})_i}$.

\begin{figure} \centering
\includegraphics*[width=2.4in]{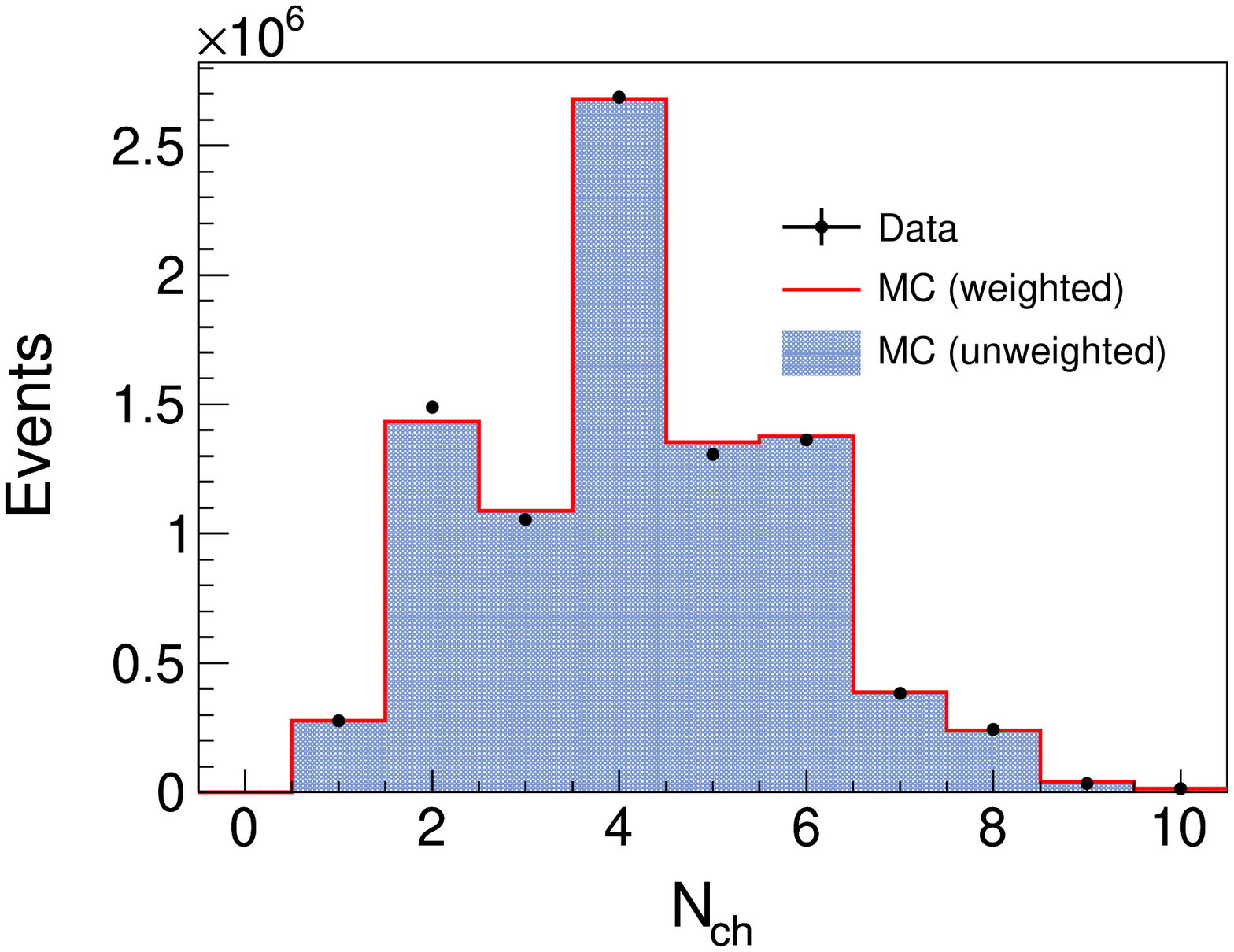}
\put(-30, 105){\bf \large  {(a)}} \\      
\includegraphics*[width=2.4in]{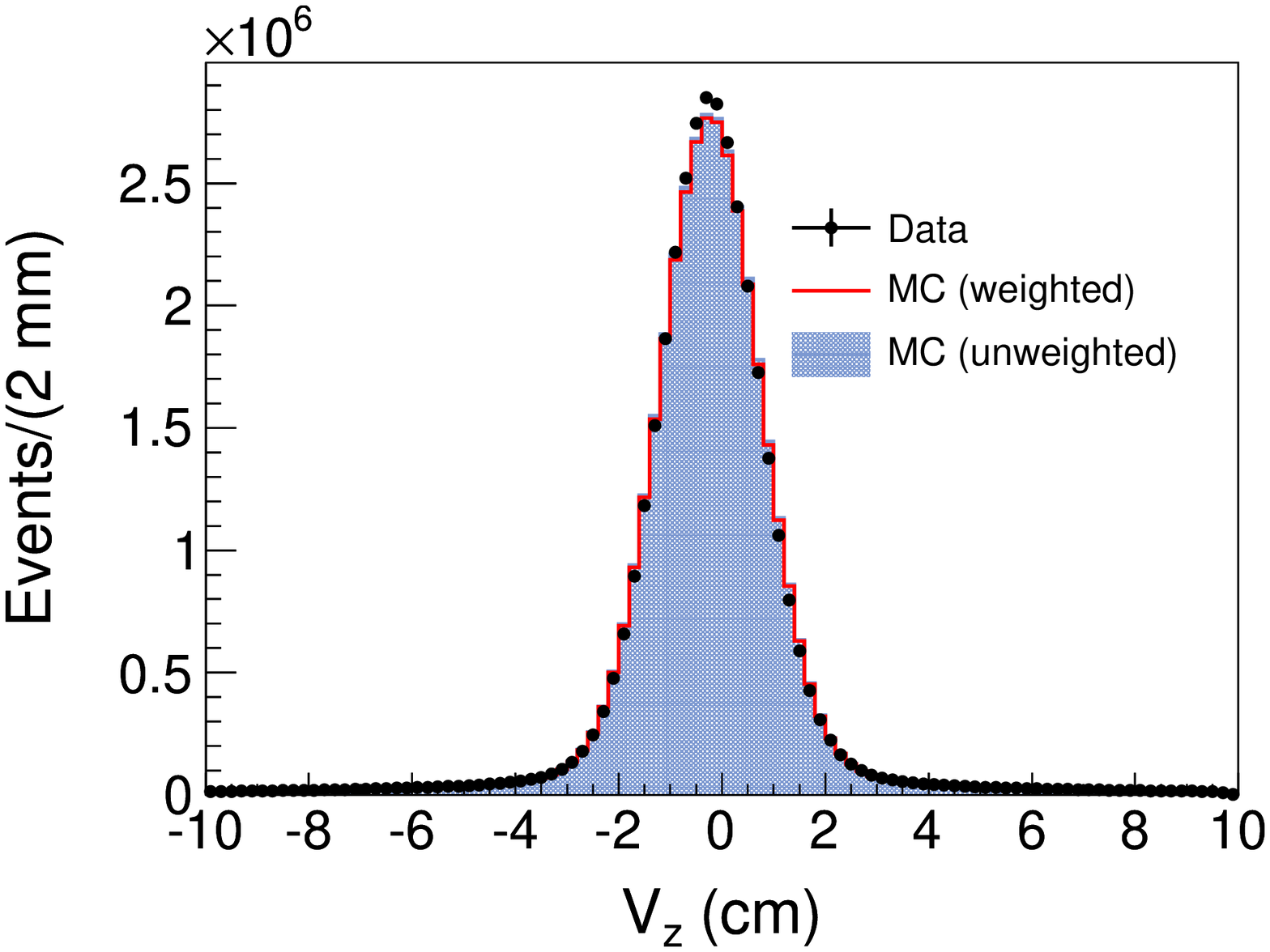}
\put(-30, 105){\bf \large  {(b)}} \\      
\includegraphics*[width=2.4in]{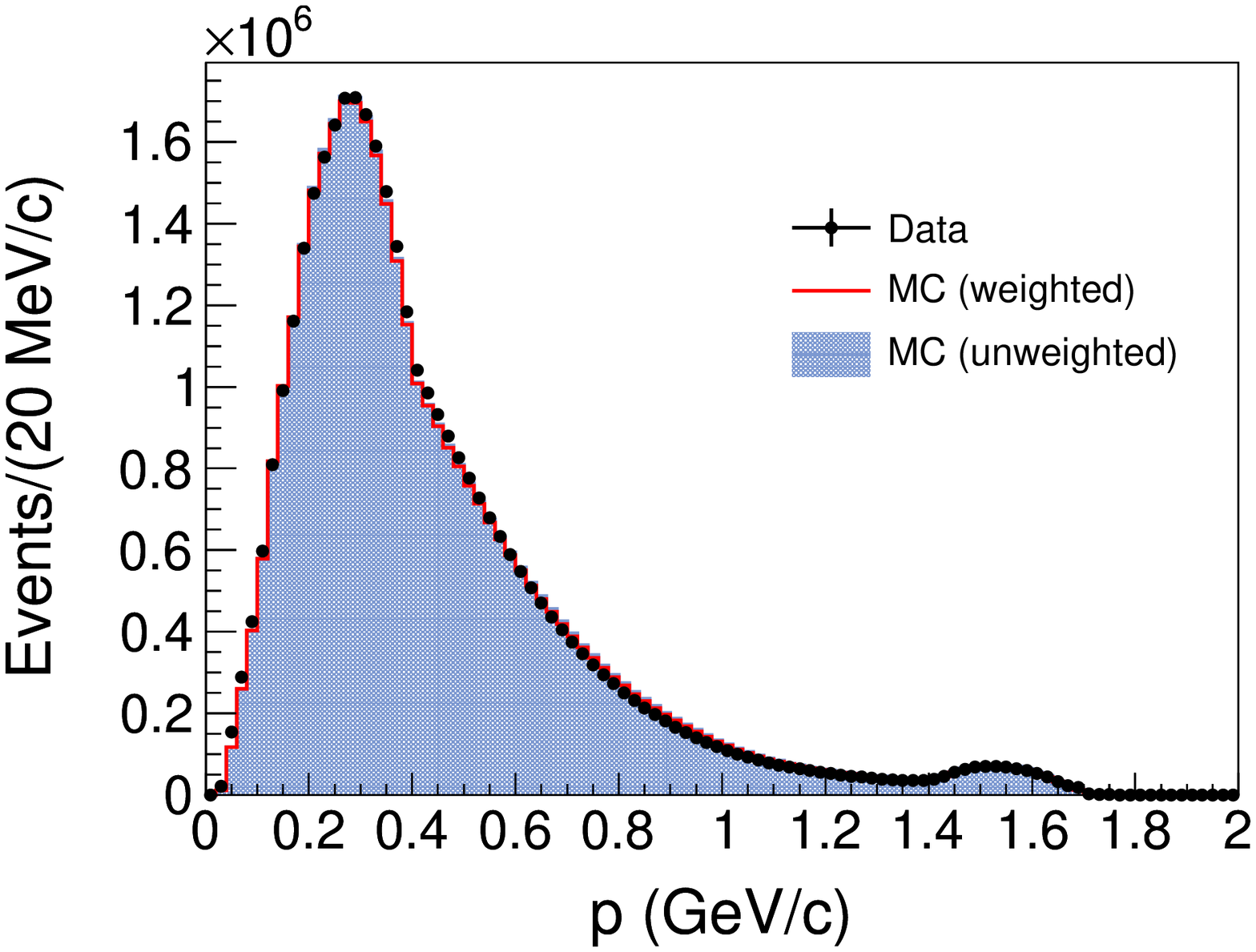}  
\put(-30, 105){\bf \large  {(c)}} \\     
\includegraphics*[width=2.4in]{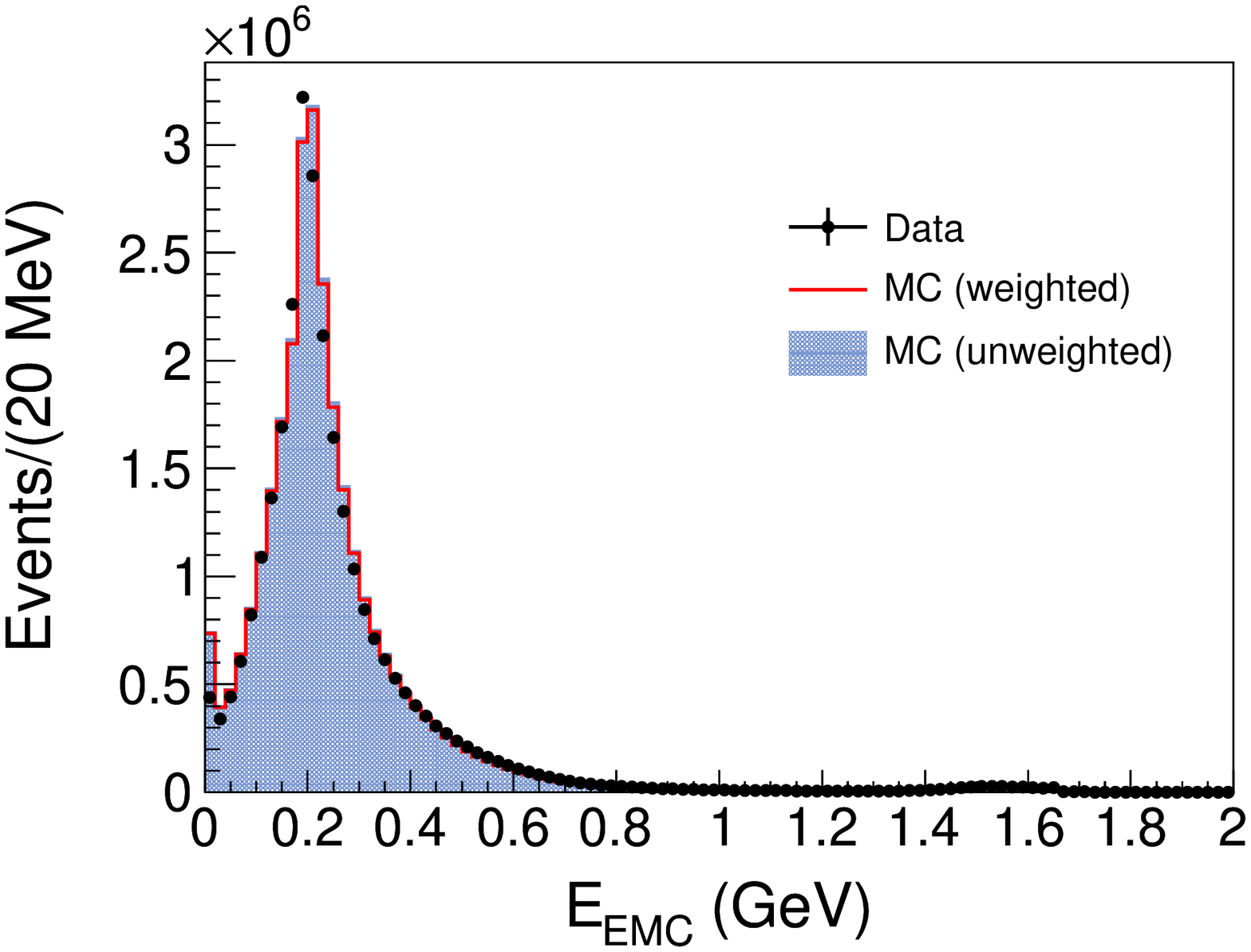}  
\put(-30, 105){\bf \large  {(d)}}      
\caption{\label{charged1} The distributions are (a) $N_{\text{ch}}$, (b)
  $V_z$, (c) $p$, and (d) $E_{\text{EMC}}$.  Data are represented by dots,
  and the MC sample by the red and shaded histograms for the weighted and
  unweighted MC events, respectively.}
\end{figure}

\begin{figure} \centering
\includegraphics*[width=2.4in]{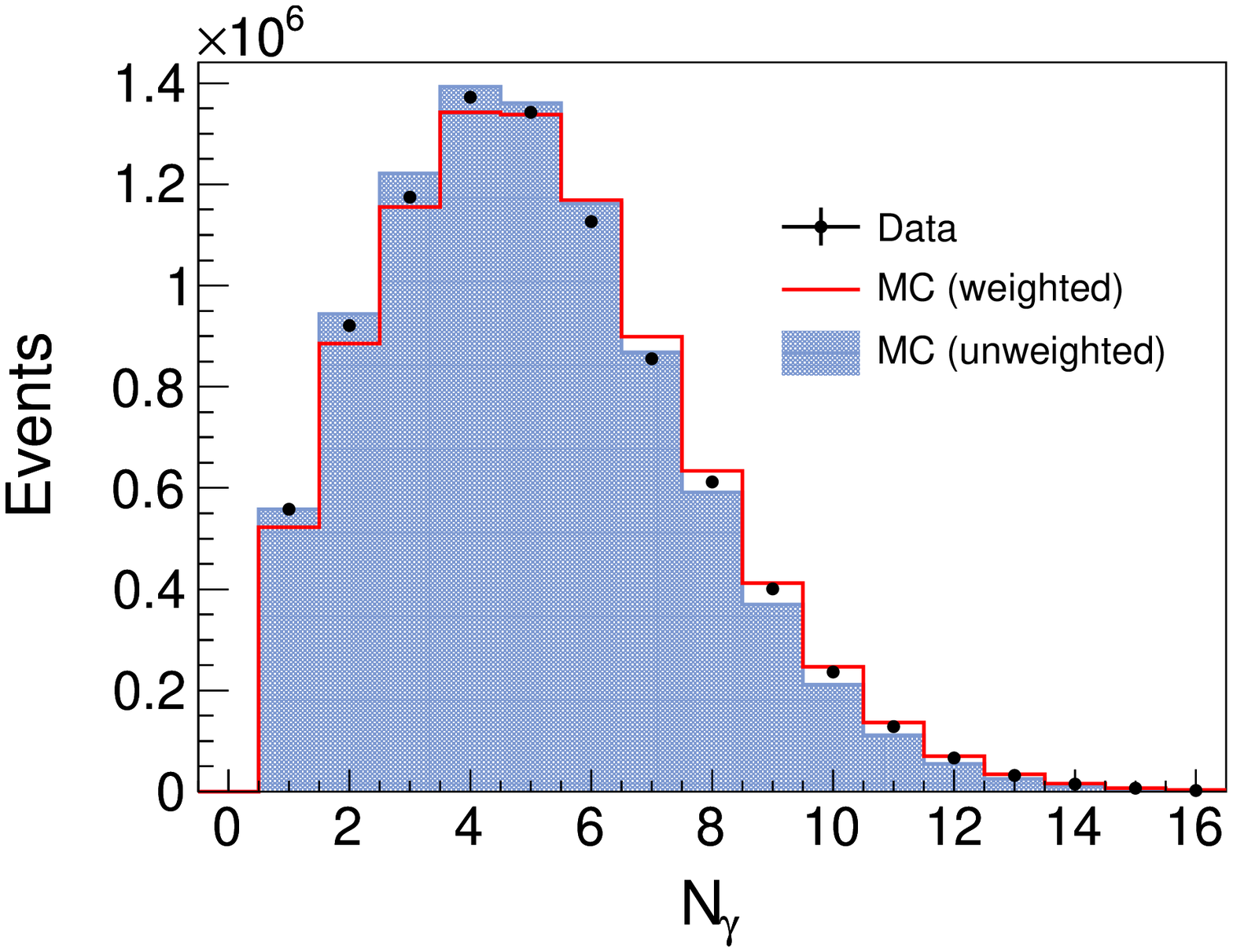}
\put(-30, 105){\bf \large  {(a)}} \\
\includegraphics*[width=2.4in]{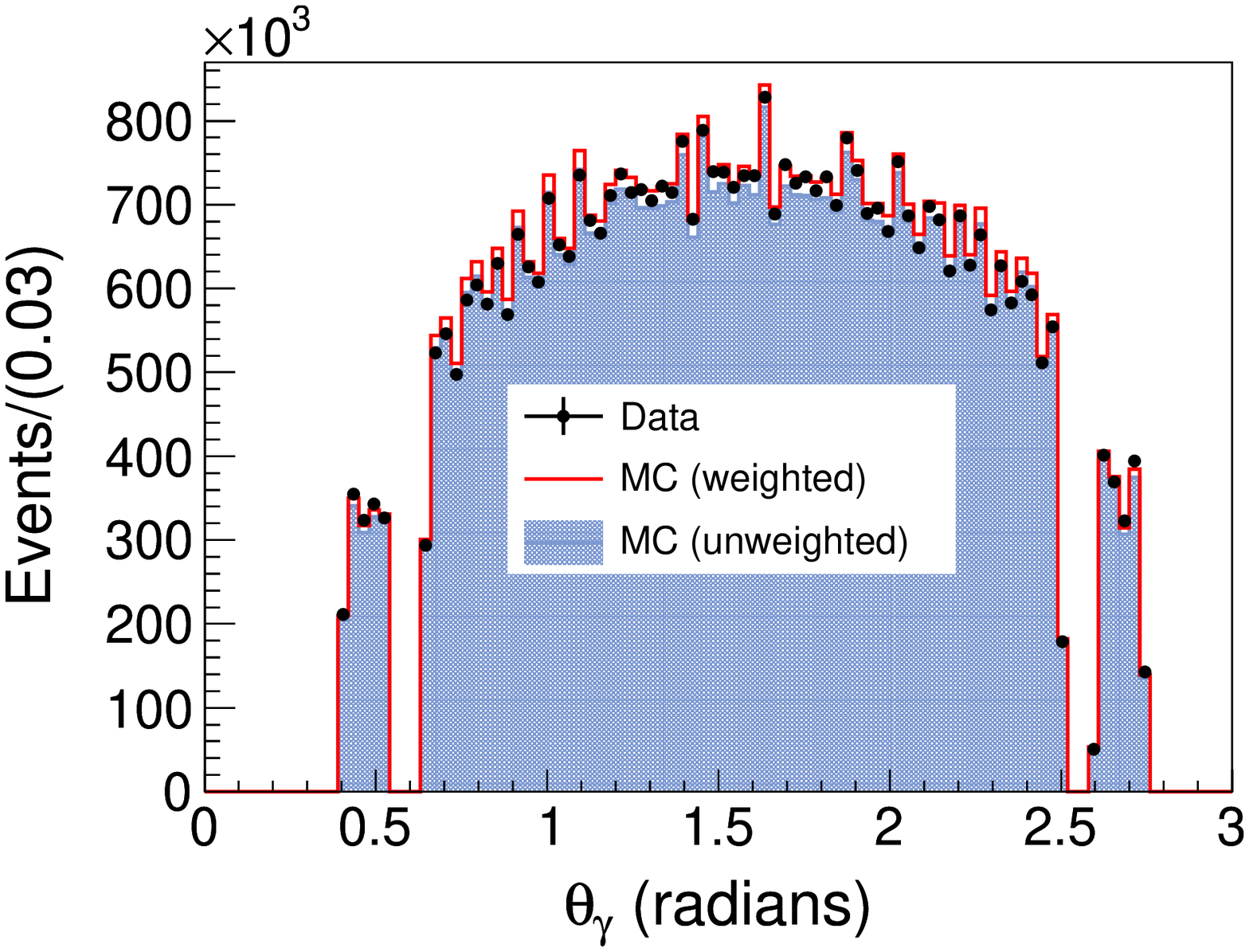}  
\put(-30, 105){\bf \large  {(b)}} \\      
\includegraphics*[width=2.4in]{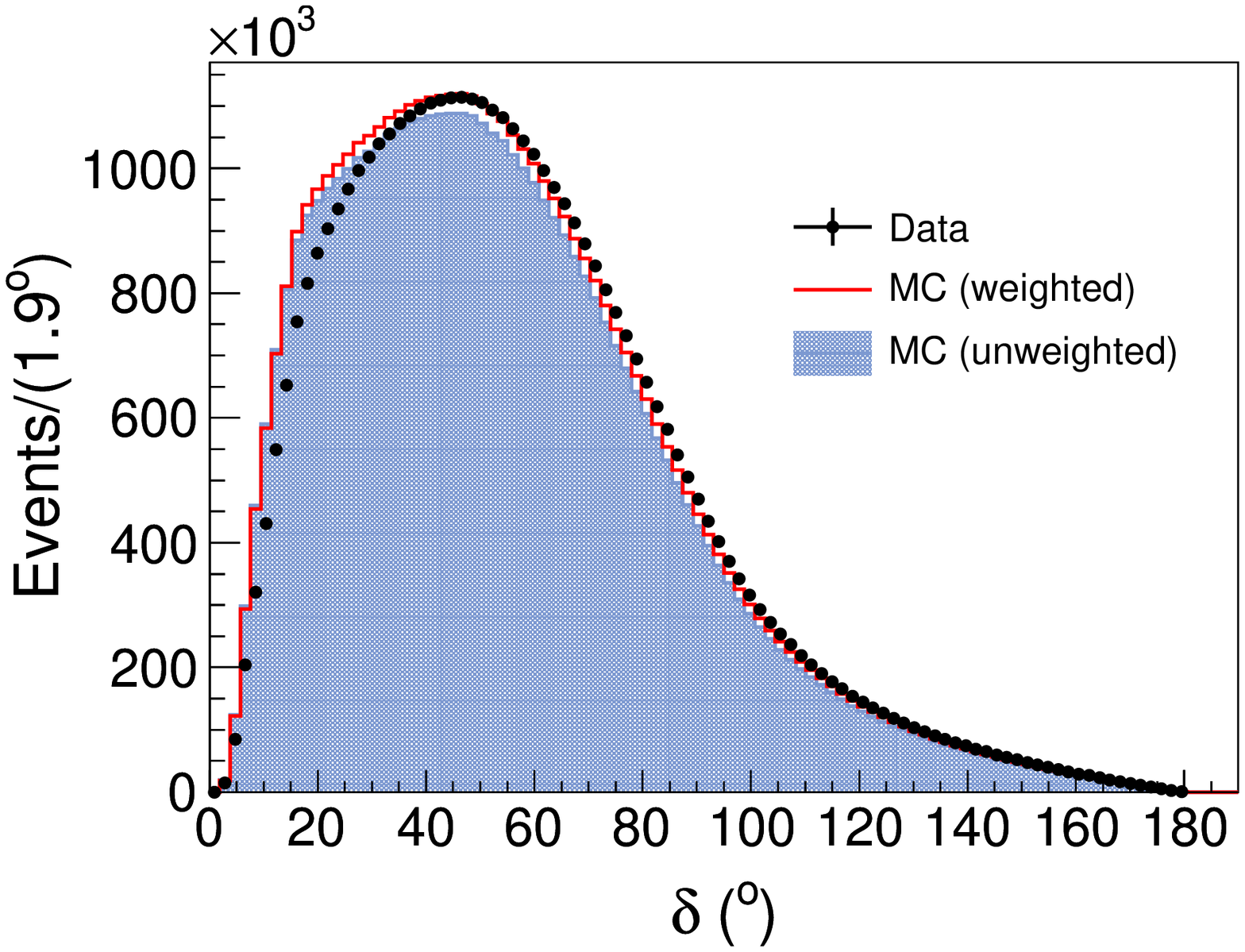} 
\put(-30, 105){\bf \large  {(c)}} \\    
\includegraphics*[width=2.4in]{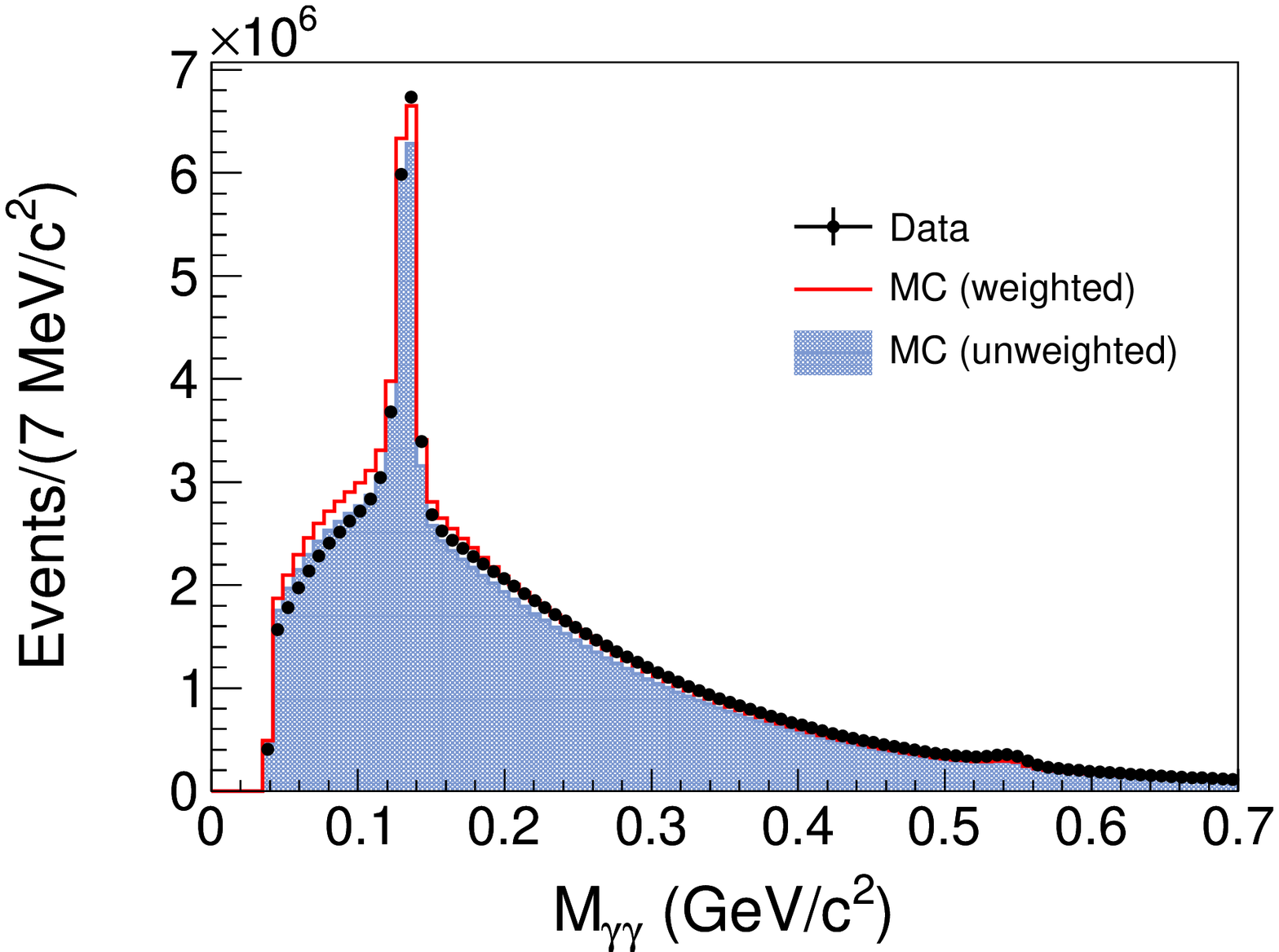}  
\put(-30, 105){\bf \large  {(d)}}      
\caption{\label{gamma} The distributions are (a) $N_{\gamma}$, (b)
  $\theta_{\gamma}$, (c) $\delta$, and (d) $M_{\gamma \gamma}$ of all
  $\gamma \gamma$ combinations. Here $\theta_{\gamma}$ is the polar
  angle of the photon.  Data are represented by dots,
  and the MC sample by the red and shaded histograms for the weighted and
  unweighted MC events, respectively.}
\end{figure}

\begin{figure} \centering
\includegraphics*[width=2.4in]{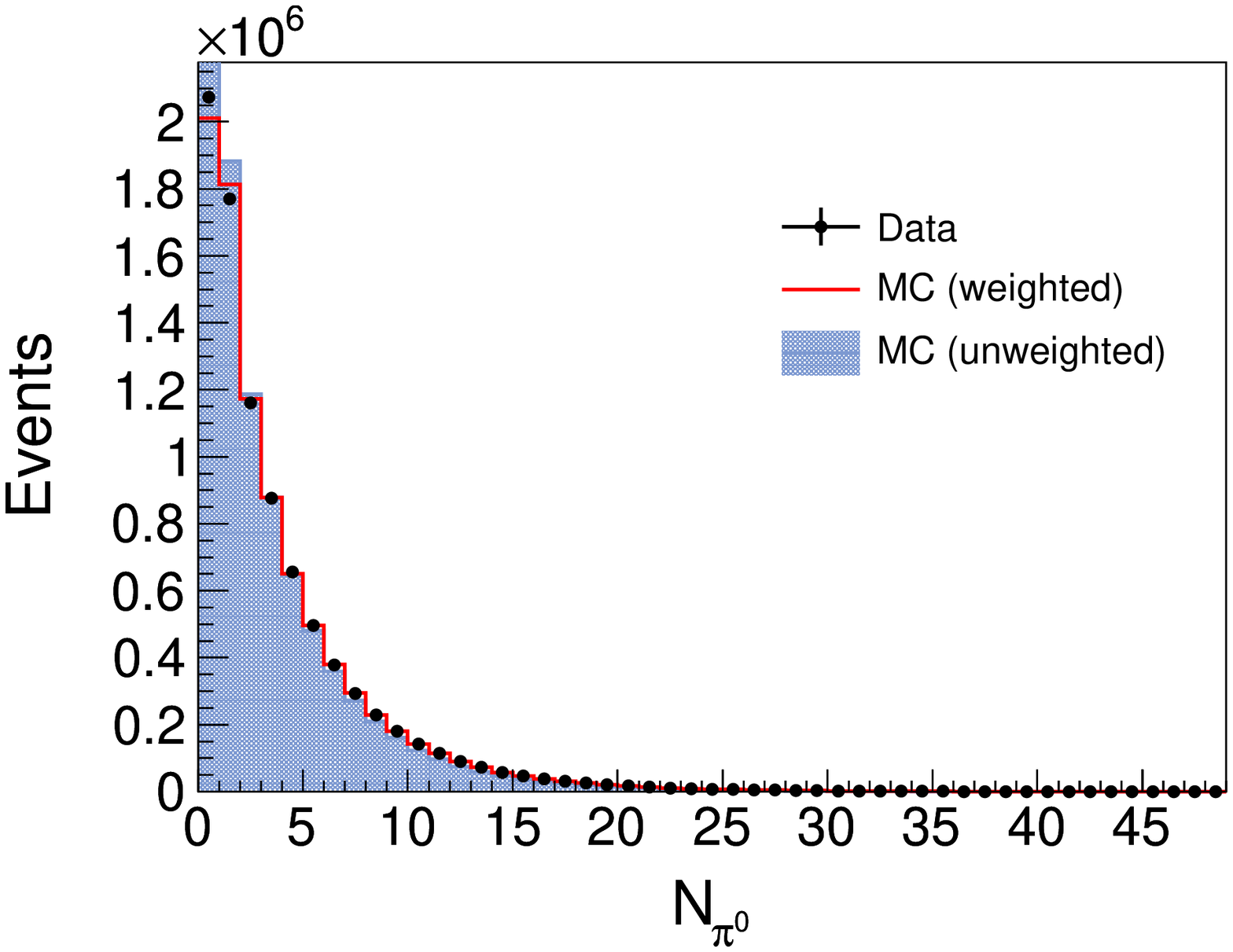}
\put(-30, 105){\bf \large  {(a)}} \\     
\includegraphics*[width=2.4in]{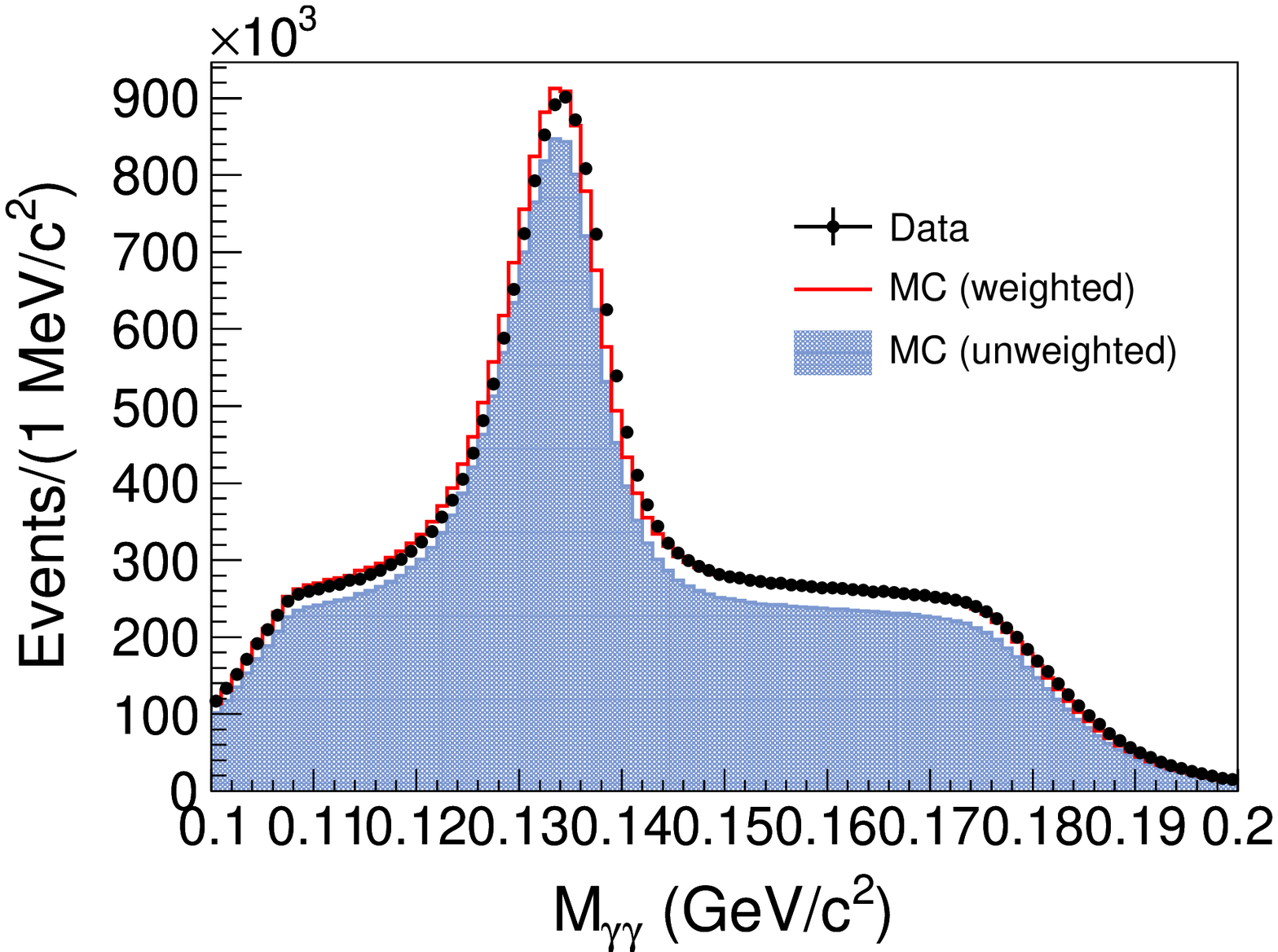}  
\put(-30, 105){\bf \large  {(b)}} \\      
\includegraphics*[width=2.4in]{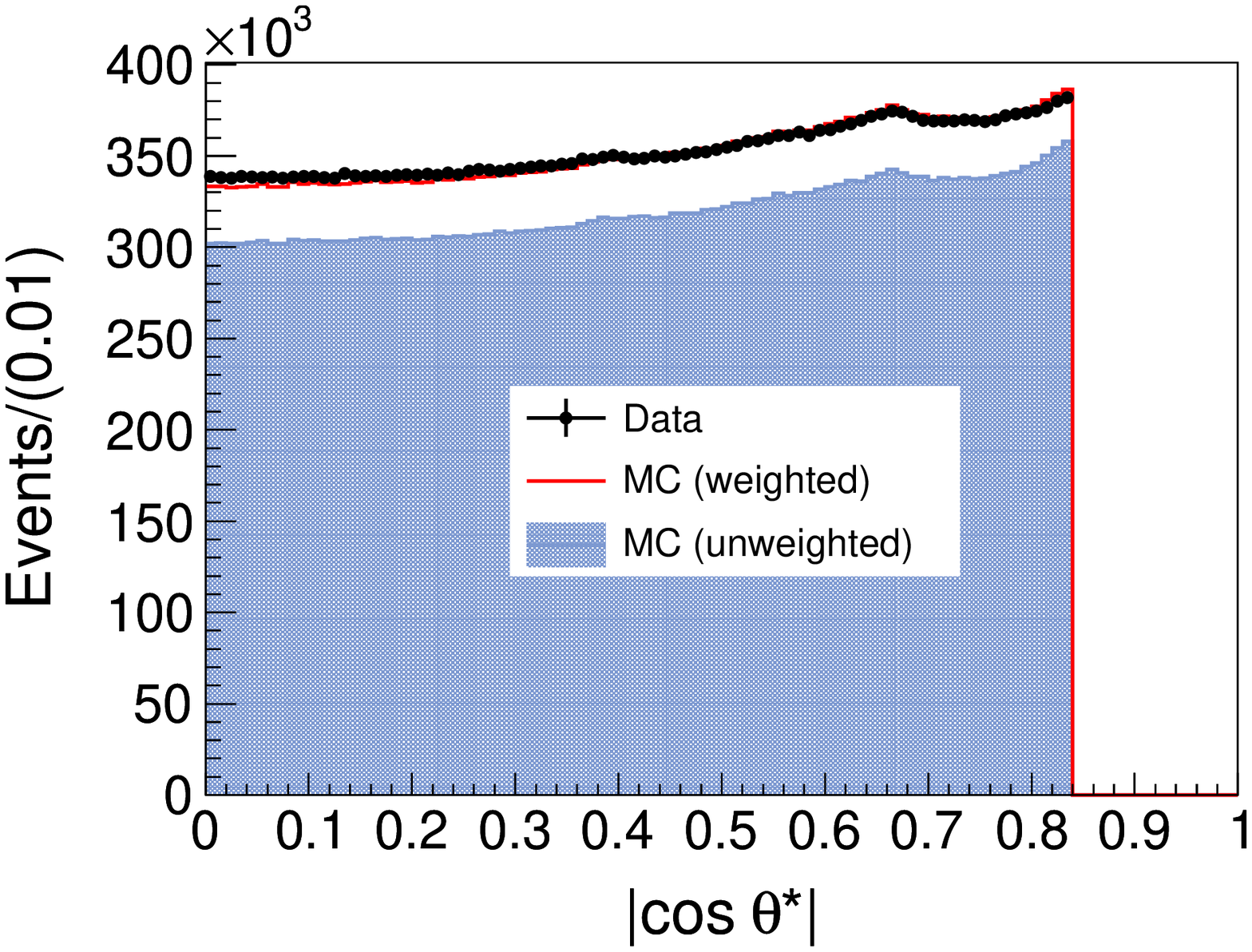}
\put(-30, 105){\bf \large  {(c)}} \\     
\includegraphics*[width=2.4in]{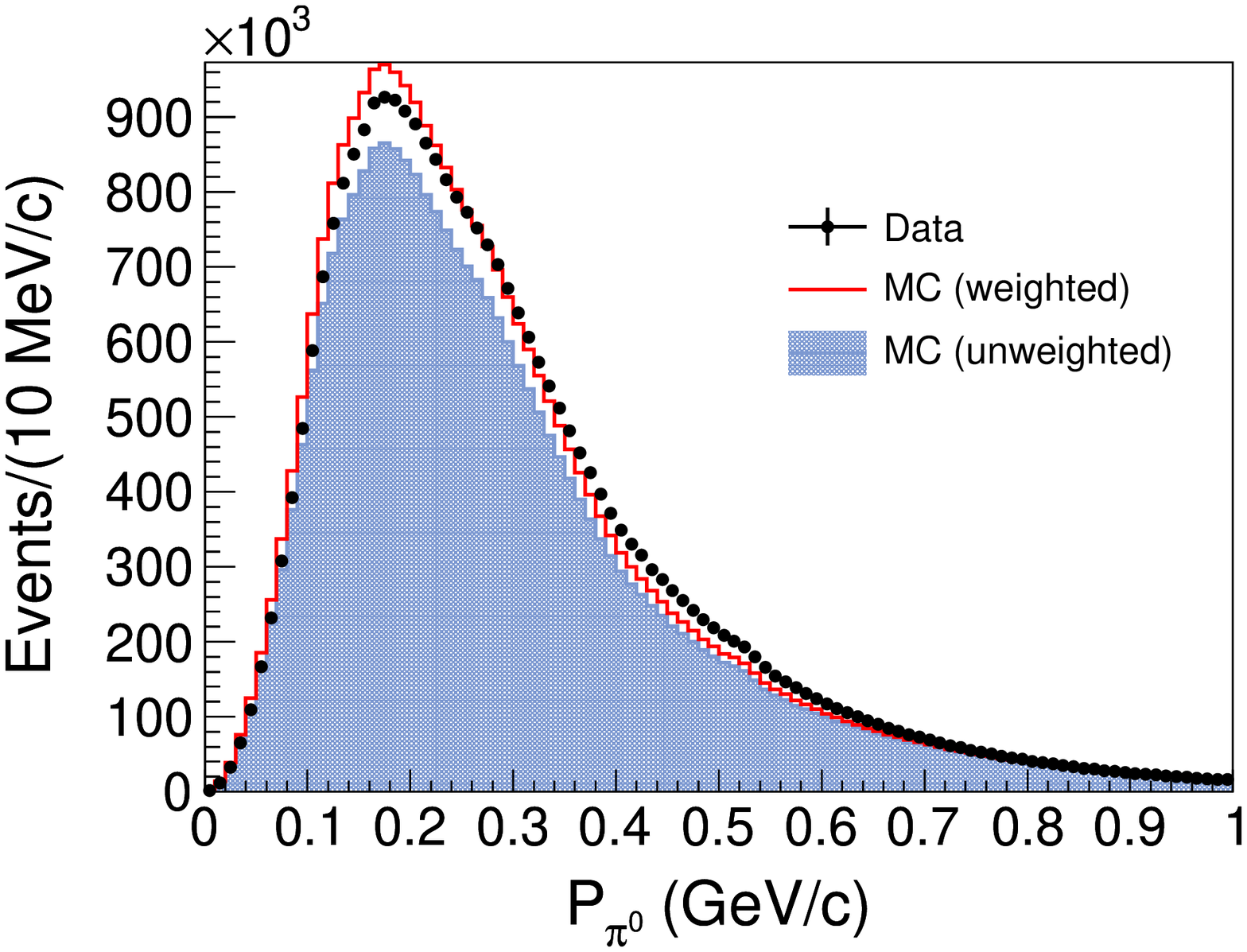}  
\put(-30, 105){\bf \large  {(d)}}      
\caption{\label{pi0}The distributions of $\pi^0$ candidates are (a)
  $N_{\pi^0}$, (b) $M_{\gamma\gamma}$ made without the $\pi^0$ mass selection requirement, (c) $|\cos \theta^*|$, (d)
  $p_{\pi^0}$.  Data are represented by dots,
  and the MC sample by the red and shaded histograms for the weighted and
  unweighted MC events, respectively. }
\end{figure}

In Fig.~\ref{charged1} representative charged track distributions, (a)
$N_{\text{ch}}$, (b) $V_z$, (c) $p$, and (d) $E_{\text{EMC}}$, are
shown.  Here and for the distributions of Figs.~\ref{gamma} and
\ref{pi0}, data, unweighted MC, and weighted MC distributions are
shown. Photon distributions, (a) $N_{\gamma}$, (b) $\theta_{\gamma}$,
(c) $\delta$, and (d) $M_{\gamma \gamma}$ of all $\gamma \gamma$
combinations, are shown in Fig.~\ref{gamma}.  The agreement is
acceptable for the charged distributions with or without
weighting. For photons, the agreement for the $\pi^0$ peak in the
$M_{\gamma \gamma}$ distribution (Fig.~\ref{gamma} (d)) is improved
with the weighted MC distribution, while the agreement for the other
distributions is neither better or worse.

Representative $\pi^0$ candidate (see Section \ref{good-pi0})
distributions, (a) the number of $\pi^0$s ($N_{\pi^0}$), (b) the
$\gamma\gamma$ invariant mass ($M_{\gamma \gamma}$) made without the
$\pi^0$ mass selection requirement, (c) $|\cos \theta^*|$, and (d)
momentum ($P_{\pi^0}$), are shown in Fig.~\ref{pi0}. The agreement is
improved for the weighted sample, and in the following, the inclusive
MC distributions will be weighted by $w_{\pi^0}$.

\section{Inclusive photon energy distributions}
\label{inclusive-section}

Inclusive photon energy distributions are obtained using the following
selection requirements.  First, the event must satisfy the inclusive
$\psi(3686)$ selection requirements, as described in
Section~\ref{inc_sel}, and not be a non-$\psi(3686)$ background event,
as defined in Section~\ref{non_psi_bkg}, a $\pi^+ \pi^- J/\psi$ event,
or a $\pi^0 \pi^0 J/\psi$ event. The $\pi^+ \pi^- J/\psi$ events are
selected with the following requirements.  There are two oppositely
charged pions with momenta $p_{\pi} < 0.45$ GeV/$c$, and the mass
recoiling from the $\pi^+ \pi^-$ system, $RM^{+-}$, must satisfy $
3.09 < RM^{+-} < 3.11$ GeV/$c^2$. The $\pi^0 \pi^0 J/\psi$ events must
have two $\pi^0$s with $p_{\pi} < 0.45$ GeV/$c$, and the mass
recoiling from the $\pi^0 \pi^0$ system, $RM^{00}$, must satisfy $
3.085 < RM^{00} < 3.12$ GeV/$c^2$.

\begin{figure} \centering
\includegraphics*[width=2.4in]{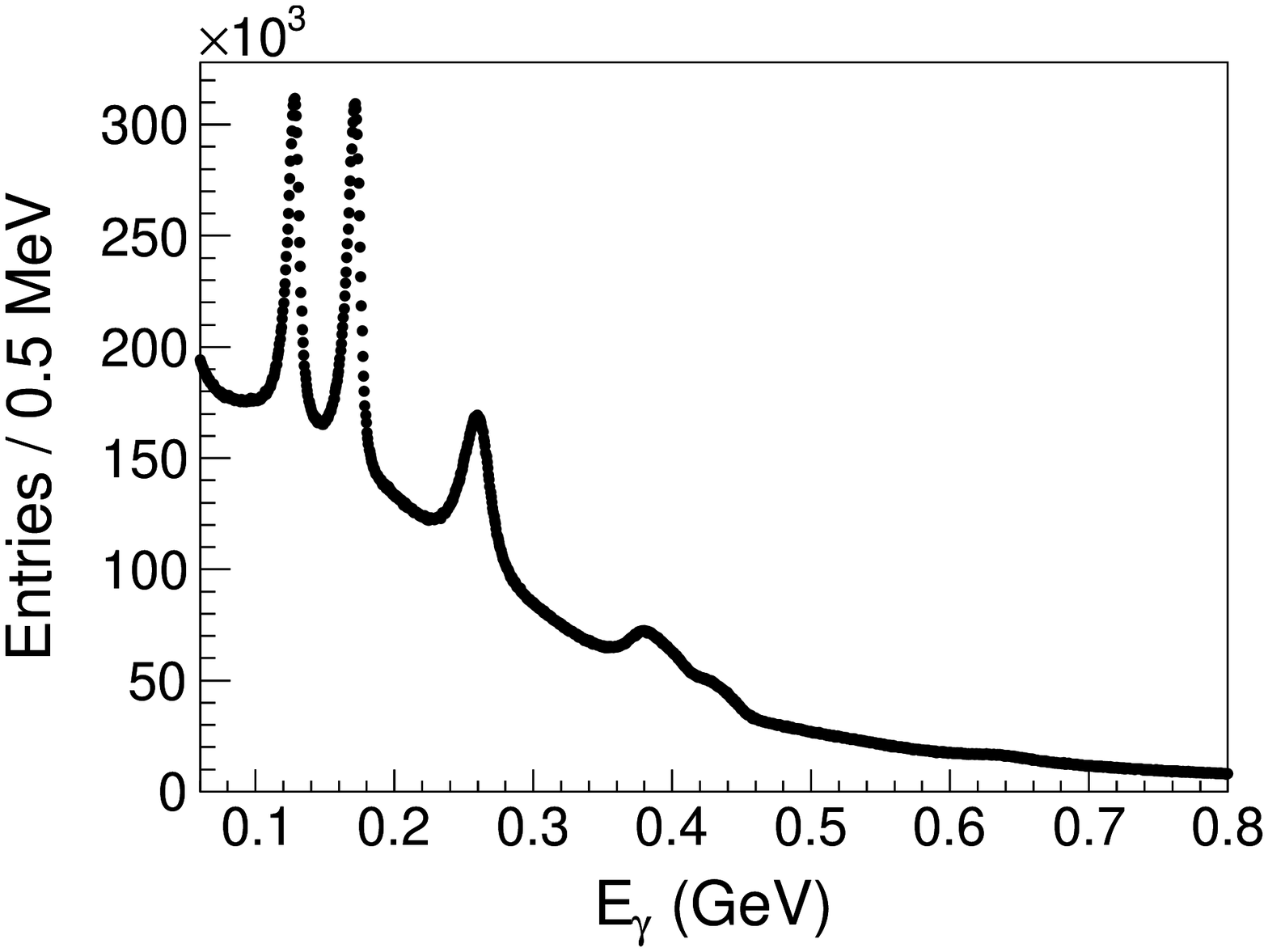}
\put(-30, 105){\bf \large  {(a)}}\\      
\includegraphics*[width=2.4in]{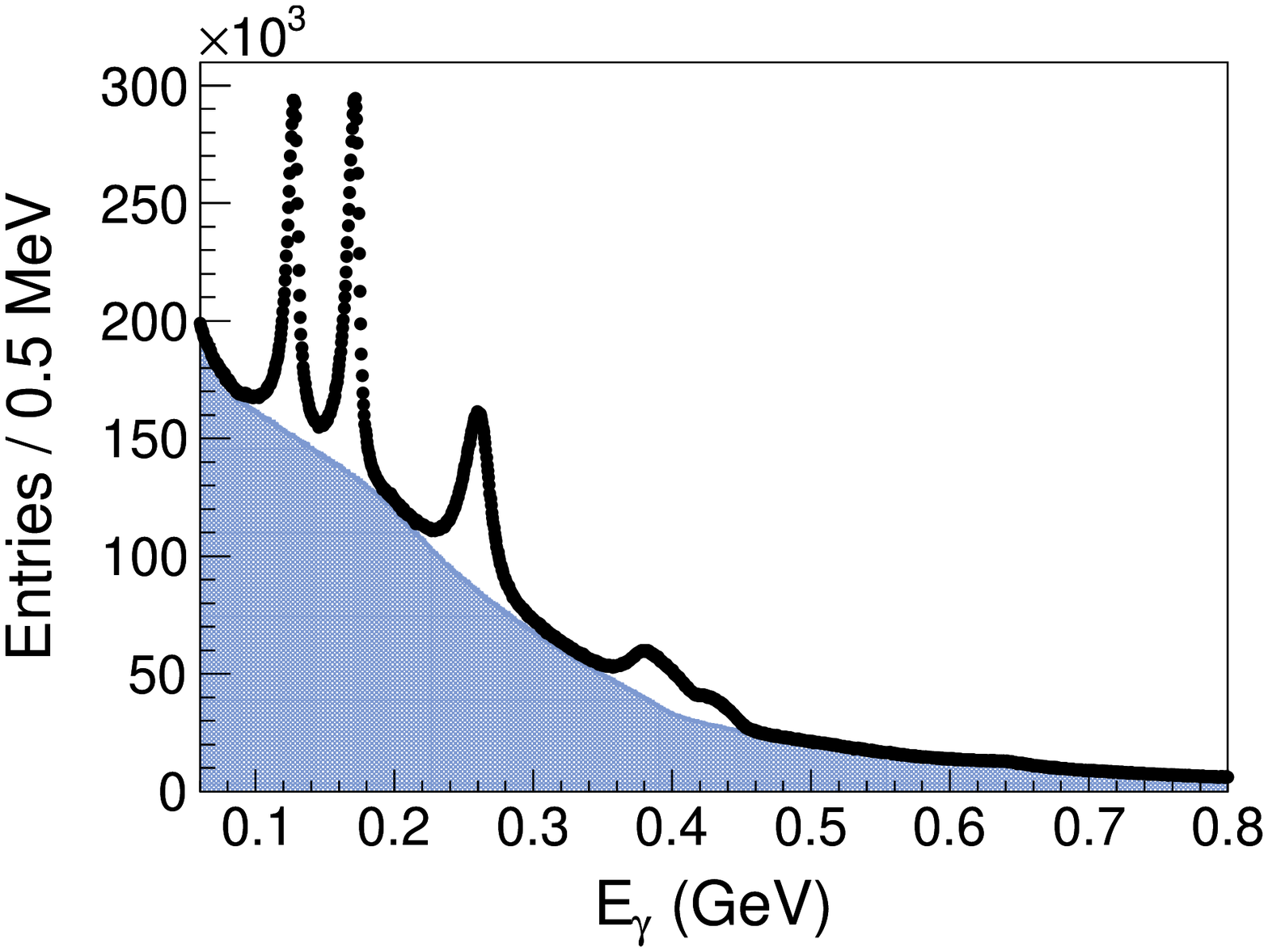}  
\put(-30, 105){\bf \large  {(b)}}      
\caption{\label{inclusive} Inclusive photon energy distributions for
  (a) data and (b) inclusive MC events, where the shaded region in (b)
  has the radiative photons removed. Peaks from left to right are
  $\psi(3686) \to \gamma \chi_{c2}$, $\gamma \chi_{c1}$, $\gamma
  \chi_{c0}$ and $\chi_{c1}$ and $\chi_{c2} \to \gamma J/\psi$. The
  $\chi_{c0} \to \gamma J/\psi$ peak is not visible.  The very small peak
  around 0.65 GeV is $\psi(3686) \to \gamma \eta_c$.  }
\end{figure}

The photon must be in the EMC barrel.  This requirement is used
because the energy resolution is better for barrel photons, and there
are fewer noise photons.  The photon must satisfy the requirement of
$\delta < 14 $ degrees (see Section~\ref{good-pi0}) and not be part of
a $\pi^0$ candidate.  In Fig.~\ref{inclusive} (a) and (b), inclusive
photon energy distributions after the above selection requirements are
shown for data and inclusive MC events, respectively.  The peaks from
left to right in each distribution correspond to $\psi(3686) \to
\gamma \chi_{c2},\; \gamma \chi_{c1},\; \gamma \chi_{c0},\; \chi_{c1}
\to \gamma J/\psi$, and $\chi_{c2} \to \gamma J/\psi$. The very small
peak at around 0.65 GeV is from the $\psi(3686) \to \gamma \eta_c$
transition.  Other small peaks not seen in the spectra but considered
in the fit are $J/\psi \to \gamma \eta_c$ and $\chi_{c0} \to \gamma
J/\psi$.

\begin{figure} \centering
\includegraphics*[width=2.4in]{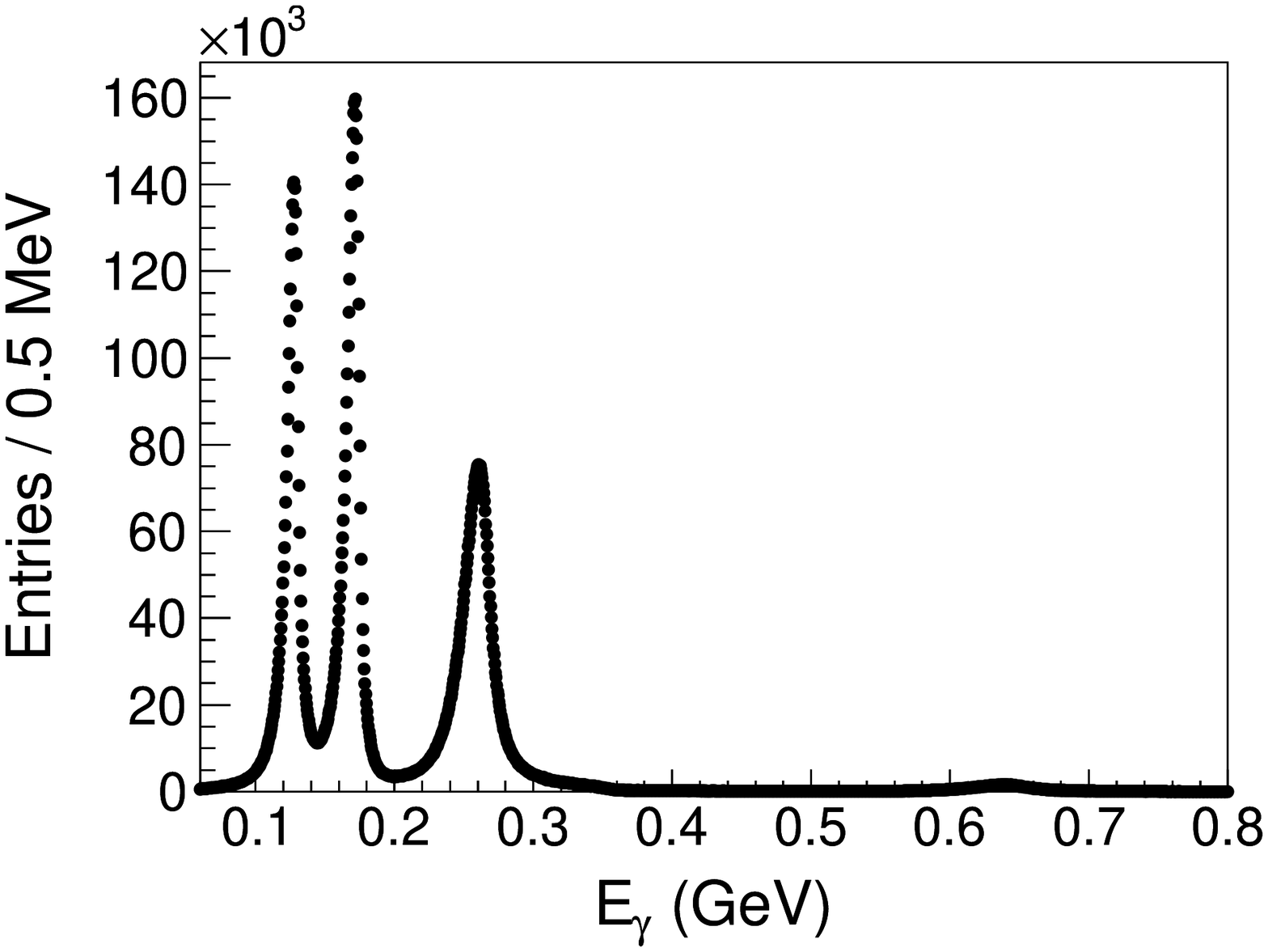}  
\put(-30, 105){\bf \large  {(a)}} \\     
\includegraphics*[width=2.4in]{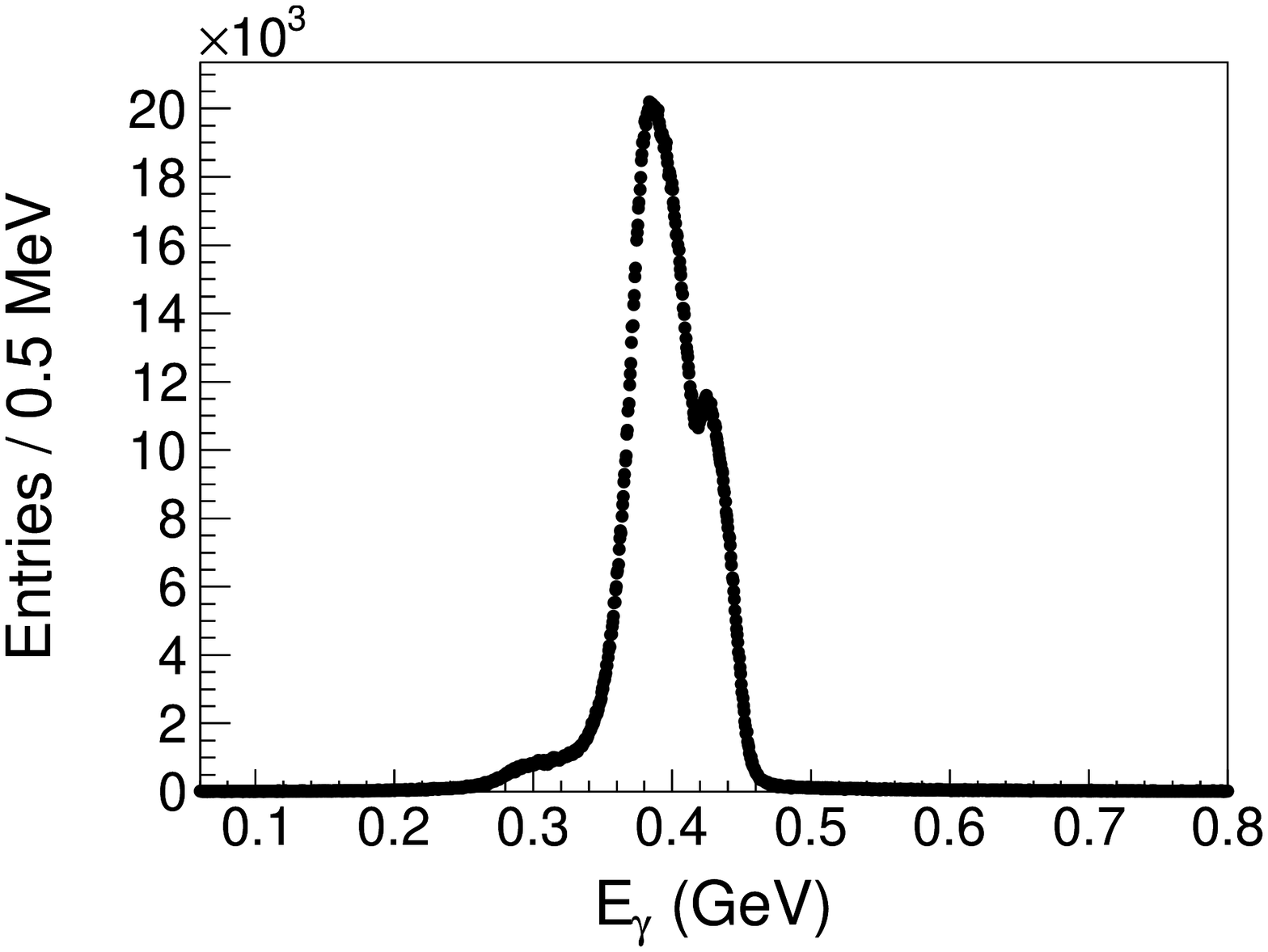}  
\put(-30, 105){\bf \large  {(b)}}      
\caption{\label{MCinclusive} Photon energy line shapes from
  inclusive MC. (a) Peaks from left to right are $\psi(3686) \to
  \gamma \chi_{c2}$, $\gamma \chi_{c1}$, and $\gamma \chi_{c0}$.  (b) Peaks
  from left to right are $\chi_{c0}$, $\chi_{c1}$, and $\chi_{c2} \to
  \gamma J/\psi$. }
\end{figure}

The inclusive $\psi(3686)$ MC sample is used to obtain the signal
shapes for charmonium transitions and the shape of the major component
of the background under the signal peaks. The signal shape for each
transition is obtained by matching the radiative photon at the
generator level with one of the photons reconstructed in the EMC.  The
requirement, which has an efficiency greater than 99\,\%, is that the
angle between the radiative photon and the reconstructed photon in the
EMC must be less than 0.08 radians.  No requirement on the energy is
used to allow obtaining the tails of the energy distribution.  The
signal shapes are shown in Fig.~\ref{MCinclusive}.  The three large
peaks from left to right in Fig.~\ref{MCinclusive} (a) correspond to
the $\psi(3686) \to \gamma \chi_{c2},\; \gamma \chi_{c1}$, and $\gamma
\chi_{c0}$ transitions. The very small peak around 0.65 GeV is the
$\psi(3686) \to \gamma \eta_c$ transition. The peaks in
Fig.~\ref{MCinclusive} (b) from left to right correspond to the
$\chi_{cJ} \to \gamma J/\psi$ transitions for $J$ = 0, 1, and 2, where
the $\chi_{c0} \to \gamma J/\psi$ peak at around 0.3 GeV is very
small.

The background component is obtained from the simulated inclusive photon energy
distribution after all selection requirements but with energy deposits
from radiative photons for charmonium radiative transition events
($\psi(3686) \to \gamma \chi_{cJ}$, $\psi(3686) \to \gamma \eta_{c}$,
$\chi_{cJ} \to \gamma J/\psi$, and $J/\psi \to \gamma \eta_c$)
removed.  Note that this distribution, shown as the shaded region in
Fig.~\ref{inclusive} (b), has a complicated shape.  This distribution
will be used to describe part of the background under the signal peaks
in fitting the data and MC inclusive photon energy distributions, as
described in Section~\ref{fitting}.

The $E1$ transition is expected to have an energy dependence of
$E_{\gamma}^3$, where $E_{\gamma}$ is the energy of the radiative
photon in the CM of the parent particle~\cite{E1}.  To account for
the $E1$ transitions for $\psi(3686) \to \gamma \chi_{cJ}, \chi_{cJ}
\to \gamma J/\psi$, a weight ($w_{\text{trans}}$) is calculated for each MC
event using the radiative photon CM energy.  For $\psi(3686) \to
\gamma_1 \chi_{cJ}$ events with no subsequent $ \chi_{cJ} \to \gamma
J/\psi$ decay, the weights are given by
$(\frac{E_{\gamma_1}}{E_{\gamma_{10}}})^3$, where $E_{\gamma1}$ for
each decay is the radiative photon CM energy and
$E_{\gamma10}$ is the most probable transition energy ($E_{\gamma10} =
\frac{E_{\text{cm}}^2-M_{\chi_{cJ}}^2}{2 \times E_{\text{cm}}})$.  For $\psi(3686)
\to \gamma_1 \chi_{cJ}, \chi_{cJ} \to \gamma_2 J/\psi$ events, the
weights are calculated according to
$(\frac{E_{\gamma_1}}{E_{\gamma_{10}}})^3
(\frac{E_{\gamma2}}{E_{\gamma_{20}}})^3$, where $E_{\gamma2}$ is the
energy of the daughter radiative photon in the rest
frame of the mother particle and $E_{\gamma_{20}}$ is its most
probable energy.  The overall event weight is the product of both
weights ($w_{\pi^0} \times w_{\text{trans}}$).

\section{\boldmath $\psi(3686) \to \gamma \chi_{cJ}$ exclusive event
  selection and photon energy distributions}

In order to constrain the final $\psi(3686) \to \gamma \chi_{cJ}$ signal
shapes in fitting inclusive photon energy distributions, clean energy
spectra from $\psi(3686) \to \gamma \chi_{cJ}, \chi_{cJ} \to$
exclusive events will be used. To fit the $\psi(3686) \to \gamma
\chi_{cJ}$ peaks of data, exclusive event samples are selected from
$\psi(3686)$ data. To fit the MC $\psi(3686) \to \gamma
\chi_{cJ}$ peaks, exclusive samples are generated, as described
below.  Exclusive events must satisfy the same requirements as
inclusive events when constructing photon energy distributions.

\subsection{\boldmath $\psi(3686) \to \gamma \chi_{cJ}$ exclusive event
  selection}
\label{chicj-exclusive}
The exclusive $\psi(3686) \to \gamma \chi_{cJ}$ photon energy
distribution is the sum of $\psi(3686) \to \gamma \chi_{cJ}, \chi_{cJ}
\to$ 2 and 4 charged track events.

\paragraph*{\bf Common requirements}

The number of good photons must be greater than zero and less that 17.
The photon with the minimum $\theta_{recoil}$, which is the angle
between the photon and the momentum recoiling from the two (four)
charged tracks, is selected as the radiative photon, and
$\theta_{recoil}$ must satisfy $\theta_{recoil} < 0.2$ rad.  Also
required are $|\cos \theta_{rad-\gamma}| < 0.75$, where
$\theta_{rad-\gamma}$ is the polar angle of the radiative photon, and
$3.3 < M_{2 (4) \pi} < 3.62$ GeV/$c^2$, where $ M_{2 (4) \pi}$ is the
invariant mass of the two (four) charged tracks.

 \paragraph*{\bf \boldmath Specific requirements for $\psi(3686) \to
   \gamma \chi_{cJ}, \chi_{cJ} \to$ 2 charged tracks}

 We require one positively and one negatively charged track.  Particle
 identification probabilities are determined using $dE/dx$ information
 from the MDC and time of flight information from the TOF system, and
 both tracks are required to be either kaons (Prob($K$) $>$ Prob($\pi$))
 or pions (Prob($\pi$) $>$ Prob($K$)). We also require $|\cos \theta|
 < 0.85$ for both charged tracks, where $\theta$ is the polar angle,
 the momentum of each track be less than 1.4 GeV$/c$, and the momentum of
 one track be larger than 0.5 GeV$/c$.

\paragraph*{\bf \boldmath Specific
requirements for $\psi(3686) \to \gamma \chi_{cJ}, \chi_{cJ} \to$ 4
charged tracks}

We require two positive and two negative tracks and $|\Sigma p_z| <
0.04$ GeV/$c$, where $| \Sigma p_z|$ is the sum of the momenta of the
charged tracks and neutral clusters in the $z$ direction.  ISR events tend to
have large $ |\Sigma p_z|$. Also the mass recoiling from the two low
momentum tracks is required to be less than 3.05 GeV/$c^2$ to veto
$\psi(3686) \to \pi \pi J/\psi$ background.

 \subsection{\boldmath $\psi(3686) \to \gamma \chi_{cJ}$ exclusive MC
   sample}
\label{exclusiveMC}
Here, exclusive $\chi_{cJ} \to$ two and four pion and kaon events are
generated with {\sc evtgen}~\cite{evtgen}, and the generated events
are selected using the selection criteria described in
Section~\ref{chicj-exclusive}. Events are weighted by $w_{\text{trans}}$
using the generated energy of the radiative photon.

\section{Fitting the inclusive photon energy distribution}
\label{fitting}

The numbers of $\psi(3686) \to \gamma \chi_{cJ}$ events and $\chi_{cJ}
\to \gamma J/\psi$ events are obtained by fitting the inclusive photon
energy distributions for data.  The efficiencies are obtained from the
fit results for the inclusive $\psi(3686)$ MC events. 

To fit the $\psi(3686) \to \gamma \chi_{cJ}$ signal peaks of
data, the MC signal shapes, described in Section~\ref{inclusive-section},
are convolved with asymmetric Gaussians to account for the difference
in resolution between MC and data, where the parameters of the
Gaussians are determined by the fit.
The broad $\chi_{c1}$ and $\chi_{c2} \to \gamma J/\psi$ peaks are
described well by just the MC shapes.
Also included in the fit are $\chi_{c0} \to \gamma J/\psi$ and $J/\psi
\to \gamma \eta_c$.  The background distribution is the inclusive MC
photon energy distribution with energy deposits from radiative photons
removed combined with a second order Chebychev polynomial function.

To constrain further the $\psi(3686) \to \gamma \chi_{cJ}$ signal
shapes, a simultaneous fit to inclusive (see
Section~\ref{inclusive-section}) and exclusive photon energy
distributions (see Section~\ref{chicj-exclusive}) is done in the
energy range from 0.08 to 0.35 GeV. The parameters of the asymmetric
Gaussians are the same for the inclusive and exclusive fits.  However,
all signal shapes are allowed to shift independently in energy for the two
distributions.  The exclusive background distribution is determined in
a similar way as the inclusive photon background distribution but
using the exclusive event selection on the $\psi(3686)$ MC event
sample.

\begin{figure}[!htpb]
\centering
\rotatebox{-90}
{\includegraphics*[width=4.0in,height=3.3in]{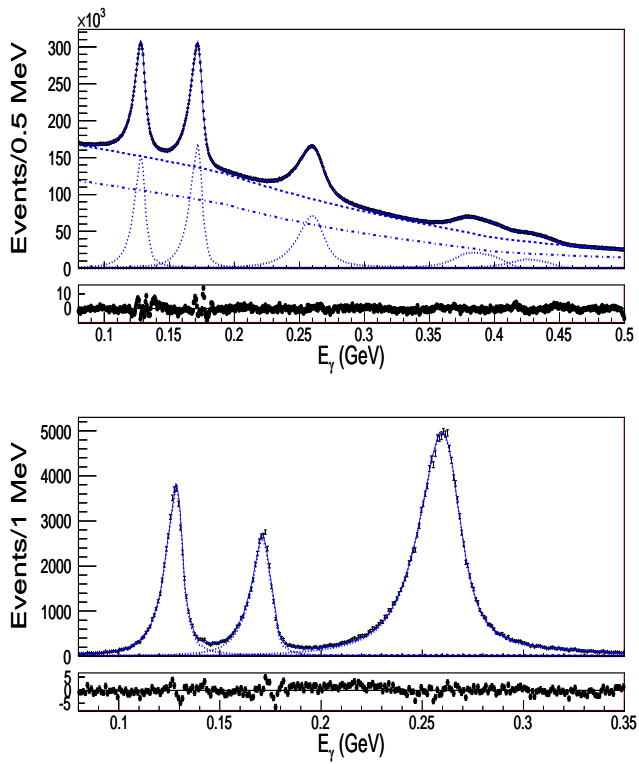}}  
\caption{\label{fit1} Simultaneous fits to the photon energy
  distributions of data. (Top set) Inclusive distribution fit and
  corresponding pulls, and (Bottom set) exclusive distribution fit and
  pull distribution.  Peaks from left to right in the top set are
  $\psi(3686) \to \gamma \chi_{c2}$, $\gamma \chi_{c1}$, and $\gamma
  \chi_{c0}$ and $\chi_{c1}$ and $\chi_{c2} \to
  \gamma J/\psi$.  The $\chi_{c0} \to \gamma J/\psi$ peak is not visible.
  The smooth curves in the two plots are the fit results. The
  dashed-dotted and dashed curves in the top plot are the background
  distribution from the inclusive $\psi(3686)$ MC with radiative
  photons removed and the total background, respectively.  The
  background in the exclusive fit plot is not visible.}
\end{figure}

Shown in Fig.~\ref{fit1} is the simultaneous fit of data for the
region $0.08 < E_{\gamma} < 0.5$ GeV for the inclusive photon energy
distribution and the region $0.08 < E_{\gamma} < 0.35$ GeV for the
exclusive photon energy distribution. The fit to the inclusive photon
energy distribution and the corresponding pull distribution are shown
in the top set of plots. The bottom set of plots are those for the
exclusive photon energy distribution.  The pull distributions are
reasonable, except in the vicinity of the $\psi(3686) \to \gamma
\chi_{c1}$ and $\gamma \chi_{c2}$ peaks.  The chi-squares per degree
of freedom ($ndf$) are 3.5 and 2.7 for the inclusive and exclusive
distribution fits, respectively.  The chi-square is determined using
$\chi^2 = \Sigma_i((n_i -n_i^f)/\sigma_i)^2$, where $n_i$, $n_i^f$,
and $\sigma_i$ are the number of data events in bin $i$, the result of
the fit at bin $i$, and the statistical uncertainty of $n_i$,
respectively, and the sum is over all histogram bins.

A fit is also done to the MC inclusive energy distribution.  The MC
shapes are used without convolved asymmetric Gaussians for the
$\psi(3686) \to \gamma \chi_{cJ}$ peaks.  Since only MC shapes are
used, it is not useful to do a simultaneous fit as there are no common
parameters to be determined in such a fit.  The fit matches the
inclusive photon energy distribution almost perfectly with a chisquare
close to zero. This is not unexpected since the signal and background
shapes come from the MC and when combined reconstruct the MC
distribution.

\section{Branching Fraction Determinations}

\begin{table*}[htb!]
\centering
\caption{Branching fraction results. The indicated
  uncertainties are statistical only.}
\vspace{0.05 in}
\begin{tabular}{l|c|c|c} \hline \hline
\T Branching Fraction &  Events ($\times 10^6$) &   Efficiency   & Branching  \\ 
      &                  &                & Fraction (\%) \B \\ \hline
\T $\mathcal{B}(\psi(3686) \to \gamma \chi_{c0})$ & $4.6871\pm0.0068$
&0.4692   & $9.389 \pm 0.014$ \\  
$\mathcal{B}(\psi(3686) \to \gamma \chi_{c1})$ & $4.9957\pm0.0054$ &
0.4740 & $9.905 \pm 0.011 $ \\  
$\mathcal{B}(\psi(3686) \to \gamma \chi_{c2})$ & $4.2021\pm0.0055$ &
0.4104 &$ 9.621 \pm 0.013 $ \B \\   \hline
\T $\mathcal{B}(\psi(3686) \to \gamma \chi_{c0})\times
\mathcal{B}(\chi_{c0} \to \gamma J/\psi)$ & $0.0123\pm0.0081$ &
0.4920 & $ 0.024 \pm 0.015 $ \\  
$\mathcal{B}(\psi(3686) \to \gamma \chi_{c1})\times
\mathcal{B}(\chi_{c1} \to \gamma J/\psi)$ & $1.8881\pm0.0053$ &
0.5155 & $ 3.442 \pm 0.010 $ \\  
$\mathcal{B}(\psi(3686) \to \gamma \chi_{c2})\times
\mathcal{B}(\chi_{c2} \to \gamma J/\psi)$ & $0.9828\pm0.0041$ &
0.5150 & $ 1.793 \pm 0.008 $  \B \\ \hline
\T $\mathcal{B}(\chi_{c0} \to \gamma J/\psi)$ & & & $0.25 \pm 0.16$ \\  
$\mathcal{B}(\chi_{c1} \to \gamma J/\psi)$ & & & $34.75 \pm 0.11$ \\  
$\mathcal{B}(\chi_{c2} \to \gamma J/\psi)$ & & & $18.64 \pm 0.08$ \B \\  
  \hline \hline
\end{tabular}
\label{fresults}
\end{table*}

The branching fractions are calculated using the
following equations
\begin{equation}
\mathcal{B}(\psi(3686) \to \gamma \chi_{cJ}) = \frac{N_{\psi(3686) \to
  \gamma \chi_{cJ}}}{\epsilon_{\psi(3686) \to \gamma
    \chi_{cJ}}\times N_{\psi(3686)}},
\end{equation}
where $\mathcal{B}(\psi(3686) \to \gamma \chi_{cJ})$ is the branching
fraction of $\psi(3686) \to \gamma \chi_{cJ}$, $N_{\psi(3686) \to \gamma
\chi_{cJ}}$ is the number of events in data from the fit,
$\epsilon_{\psi(3686) \to \gamma \chi_{cJ}}$ is the efficiency
determined from MC, and $N_{\psi(3686)}$ is the number of $\psi(3686)$
events~\cite{Npsip}.
The product branching fraction for $\psi(3686) \to \gamma \chi_{cJ},
  \chi_{cJ} \to \gamma J/\psi$ is given by
\begin{eqnarray}
\mathcal{B}(\psi(3686) \to \gamma \chi_{cJ})\times \mathcal{B}(\chi_{cJ} \to \gamma J/\psi) \nonumber \\
 = \frac{N_{\chi_{cJ} \to
  \gamma J/\psi}}{\epsilon_{\chi_{cJ} \to \gamma
    J/\psi}\times N_{\psi(3686)}},
\end{eqnarray}
where $N_{\chi_{cJ} \to \gamma J/\psi}$ is the number of $\chi_{cJ}
\to \gamma J/\psi$ events in data and $\epsilon_{\chi_{cJ} \to \gamma
  J/\psi}$ is the efficiency.  From Eq. (1) and Eq. (2), we obtain the
branching fraction for $\chi_{cJ} \to \gamma J/\psi$, which is given by
\begin{eqnarray}
\mathcal{B}(\chi_{cJ} &\to& \gamma J/\psi) \nonumber \\
& = &\frac{\mathcal{B}(\psi(3686) \to \gamma \chi_{cJ})\times \mathcal{B}(\chi_{cJ} \to \gamma J/\psi)}{\mathcal{B}(\psi(3686) \to \gamma \chi_{cJ})} \nonumber \\
          & = &   \frac{\epsilon_{\psi(3686) \to \gamma
    \chi_{cJ}}\times N_{\chi_{cJ} \to
  \gamma J/\psi}}{\epsilon_{\chi_{cJ} \to \gamma
    J/\psi}\times N_{\psi(3686) \to
  \gamma \chi_{cJ}}}.
\end{eqnarray}

Results are listed in Table~\ref{fresults}, where the uncertainties
are statistical only.  For $\mathcal{B}(\chi_{cJ} \to \gamma J/\psi)$,
an alternative parametrization in terms of $N_{\psi(3686) \to \gamma
  \chi_{cJ}}$ and the ratio $N_{\chi_{cJ} \to \gamma
  J/\psi}/N_{\psi(3686) \to \gamma \chi_{cJ}}$ has been tried because
  of the possible correlation between the numerator and denominator of
  Eq.~(3), but the difference with the original result is small and
  will be neglected since it is much less than the systematic
  uncertainties that will be discussed below.

\section{Systematic Uncertainties}

Systematic uncertainties, which arise from selection requirements,
fitting, photon efficiency, the uncertainty in the number of
$\psi(3686)$ events, etc. are summarized in Table~\ref{systematics}.
For $\psi(3686) \to \gamma \chi_{cJ}$, they are under 4\,\% and
smaller than those of CLEO~\cite{cleo04}, with the largest
contribution coming from fitting the photon energy distribution.
Details of how they are estimated are given below.

\begin{table*}[tb]
\centering
\caption{Relative systematic uncertainties (\%).  $\mathcal{B}_{\psi J}$ is
  notation for $\mathcal{B}(\psi(3686) \to \gamma \chi_{cJ})$,
  $\mathcal{B}_{PJ}$ is for $\mathcal{B}(\psi(3686) \to \gamma
  \chi_{cJ})\times \mathcal{B}(\chi_{cJ} \to \gamma J/\psi)$, and
  $\mathcal{B}_{\chi J}$ is for $\mathcal{B}(\chi_{cJ} \to \gamma
  J/\psi)$. Some uncertainties cancel in the determination of
  $\mathcal{B}_{\chi1}$ and $\mathcal{B}_{\chi2}$ and are left blank in the table. Since the fit
  uncertainty is so large for $\psi(3686) \to \gamma \chi_{c0},
  \chi_{c0} \to \gamma J/\psi$, the other systematic uncertainties for
  $\mathcal{B}_{P0}$ and $\mathcal{B}_{\chi 0}$ are omitted.}
\vspace{0.05 in}
\begin{tabular}{l|ccc|ccc|ccc} \hline \hline
 \T & ~$\mathcal{B}_{\psi0}$~ &   ~$\mathcal{B}_{\psi1}$~  &
  ~$\mathcal{B}_{\psi2}$~  &   ~$\mathcal{B}_{P0}$~ &   ~$\mathcal{B}_{P1}$~ &
  ~$\mathcal{B}_{P2}$~ & ~$\mathcal{B}_{\chi0}$~ & ~$\mathcal{B}_{\chi1}$~ &   ~$\mathcal{B}_{\chi2}$~
 \B  \\  \hline
  \T $N_{\text{ch}} > 0$   & 0.74 & 0.27 & 0.75  &  & 0.06 & 0.74 &  & 0.21 & 1.5\\
  $N_{\gamma} < 17$ & - &- &- &  & - &- & & & \\
  $E_{\text{vis}} > 0.22E_{\text{cm}}$ & 0.17 &0.17 &0.17 &   &0.17 &0.17 & & & \\
  $\delta > 14^{\circ}$  & 0.36 & 0.14& 0.00&   & 0.02 &1.56 &  & 0.12&1.42\\
  $\psi(3686)$ background veto &0.51 &0.73 &0.15 &  & 0.51 &0.11 &  & 1.25 & 0.26\\
  $\pi^+ \pi^- J/\psi$ veto  & 0.25 & 0.25 & 0.25 &  & 0.25 & 0.25 & &  &\\
  $\pi^0 \pi^0 J/\psi$ veto & 0.03 &0.03 &0.03 &   &0.03 &0.03 & & &\\
  $\gamma$ not in $\pi^0$ & 0.87& 0.53& 0.19&  & 1.24 & 2.3 &  &1.35 & 2.3 \\
  Fitting & 2.62 &2.69 &1.5 & 869 &3.10 &7.22 & 861 & 4.43& 7.27\\
  MC signal shape  & 0.06 & 0.17 & 0.53 &  & 0.07 & 0.53 &
   &  0.24 & 1.05 \\
Multipole correction &  0.0 & 0.61 & 0.60 & &0.35 &3.82 & &0.70 & 3.87  \\  
  $|\cos \theta| < 0.8$ & 0.49 & 0.12  & 0.07 &  & 0.35 & 1.46 &  &0.47 & 1.52 \\
 $\pi^0$ weight &1.19 &1.55 & 1.60 &  &1.09 &1.73 &  &0.47 &
  0.13\\ 
Continuum energy difference & 0.75 & 0.06 & 0.43 & & 0.35 & 0.60 &
& 0.39 & 1.02\\ 
  $\gamma$ efficiency &1.0 &1.0 &1.0 &  & 1.0& 1.0& & & \\
  $N_{\psi(3686)}$ &0.81 &0.81 &0.81 &  & 0.81&0.81 & & & \B \\ \hline
 \T Total &3.54 &3.57 &2.83 & 869 & 3.84  &9.09 & 861 & 4.92 &9.05 \B \\  
  \hline \hline
\end{tabular}
\label{systematics}
\end{table*}

\subsection{\boldmath Systematic uncertainties from initial $\psi(3686)$ event selection}
\label{systematic1}
Initial $\psi(3686)$ event selection requirements are $N_{\text{ch}} > 0$,
$N_{\gamma} < 17$, and $E_{\text{vis}} > 0.22  E_{\text{cm}}$.
To determine the systematic uncertainties associated with the $N_{\text{ch}} > 0$
requirement, events without charged tracks are also analyzed.  The
photon time requirement is removed for these events since without
charged tracks, the event start time can not be well determined. The selection
requirements are also changed because these events have much more
background.  Events must have total energy greater than 1.7 GeV and at
least one good neutral pion.  Even so, there is a background from low
energy photons, and even after subtracting continuum, the
photon energy distribution for data has a large background under the
signal peaks, making fits difficult with the number of fitted events
having large uncertainties.

The photon energy distributions for data and MC are fitted.  The
numbers of fitted events for data and MC are then added with the
number of fitted events with charged tracks, and the branching fractions
are recalculated.  The differences with the branching fractions
determined with charged track events only are then determined and
taken as the systematic uncertainties associated with the
$N_{\text{ch}} > 0$ requirement.

As described in Section~\ref{compare}, inclusive $\psi(3686)$ MC events
are weighted according to the the $N_{\pi^0}$ distribution to give
better agreement with data.  According to the MC, the
efficiency of the $N_{\gamma} < 17$ requirement, defined as the number
of events with $N_{\gamma} < 17$ divided by the number of events with
$N_{\gamma} >0$, is 99.99\,\% with weighting and 99.99\,\%
without weighting, while the efficiency for data is 99.98\,\%.  The
agreement is excellent, the efficiency is very high, and the
systematic uncertainty is negligible for this requirement.


The agreement between the $E_{\text{vis}}$ distribution of data and the
inclusive $\psi(3686)$ MC distribution is very good. 
According to the inclusive MC, the efficiency of the $E_{\text{vis}} >
0.22 E_{\text{cm}}$ requirement after the $N_{\text{ch}}$ and
$N_{\gamma}$ requirements is 99.76\,\%.  The mean and
root-mean-squared values of the MC (data) are 3.004 (2.991) GeV and
0.561 (0.579) GeV, respectively. If the MC distribution is shifted
down by 13 MeV relative to the data, the loss of events due to the
$E_{\text{vis}}$ requirement corresponds to an inefficiency of
0.17\,\%, and this will be taken as the systematic uncertainty for the
$E_{\text{vis}}$ requirement.

\subsection{\boldmath Systematic uncertainties from inclusive photon selection}

Further selection criteria are used before including photons into the
photon energy distributions which are used for fitting.  Photon selection
requirements include $\delta > 14^o$, removal of non-$\psi(3686)$ background
events, removal of $\pi \pi J/\psi$ events, and removal of photons which
can be part of a $\pi^0$. 

\paragraph*{\bf \boldmath $\delta > 14^o$ and $\psi(3686)$ background removal systematic uncertainties:\\}
To determine the systematic uncertainties for the first two
requirements, they are removed from the selection process,
and the branching fraction results obtained are compared to those with the
requirements. Removing the $\delta$ requirement changes the inclusive
photon energy background distribution of the inclusive MC, as well as the
inclusive photon energy distribution of data.  The differences of the
branching fraction results are taken as the systematic uncertainties
for each of the requirements.

\paragraph*{\bf \boldmath  $\pi^+ \pi^- J/\psi$ event removal
  systematic uncertainty:\\}

The distribution of mass recoiling from the $\pi^+\pi^-$ system,
$RM^{+-}$, for events passing the non-$\psi(3686)$ veto and the
$\psi(3686) \to \pi^+\pi^- J/\psi$ selection, but without the recoil
mass requirement in Section~\ref{inclusive-section}, has a clear
$J/\psi$ peak from $\psi(3686) \to \pi^+\pi^-J/\psi$.  Events with
$RM^{+-}$ satisfying $3.09 < RM^{+-} < 3.11$ GeV/$c^2$ will be removed
from further consideration.  However, there are $\pi^+\pi^-$
mis-combinations underneath the peak in the $J/\psi$ region. To
estimate the probability that a good radiative photon event may be
vetoed accidentally (or the efficiency with which it will pass this
veto requirement), the sideband regions, defined as $ 3.07 < RM^{+-} <
3.085$ GeV$/c^2$ and $ 3.115 < RM^{+-} < 3.13$ GeV$/c^2$, are used to
estimate the number of mis-combinations in the signal region.  Using
this veto probability, the efficiency for inclusive MC events to pass
the $\psi(3686) \to \pi^+\pi^- J/\psi$ veto requirement is found to be
93.06\,\%.  The efficiency for data is 92.83\,\%, and the difference
between data and inclusive MC is 0.23/93.06 = 0.25\,\%, which we take
as the systematic uncertainty due to the $\pi^+ \pi^- J/\psi$ veto for
all radiative photon processes.

\paragraph*{\bf \boldmath  $\pi^0 \pi^0 J/\psi$ event removal
  systematic uncertainty:\\}

The approach to determine the systematic uncertainty for the $\pi^0 \pi^0
J/\psi$ event removal is similar to that described in the previous section. 
Using the veto probability obtained using sidebands, the efficiency
for inclusive MC events to pass the $\psi(3686) \to \pi^0\pi^0 J/\psi$
veto requirement is found to be 95.34\,\%. 
The efficiency for data is 95.37\,\%, and the difference between data
and inclusive MC is 0.03/95.35 = 0.03\,\%, which we will take as the
systematic uncertainty due to the $\pi^0 \pi^0J/\psi$ veto for all
radiative photon processes. 

\paragraph*{\bf \boldmath Systematic uncertainty for the removal of
  photons which can be part of a $\pi^0$:\\}

As described in Section~\ref{good-pi0}, photons that are part of a
$\pi^0$ are excluded from the inclusive photon energy
distribution.  To estimate the systematic uncertainty for this requirement,
the efficiency of this criterion is determined for data and MC events for
each transition by fitting the photon inclusive energy distribution
with and without the $\pi^0$ removal using non-simultaneous
fitting. The systematic uncertainties are determined by the differences
between the efficiencies for data and MC events.

\subsection{Fitting systematic uncertainty}
\label{fitsys}

The systematic uncertainty associated with the fit procedure is
determined by comparing various fitting methods.  The fit is done with
an alternative strategy, fitting with a non-simultaneous fit, changing the
order of the polynomial function used from second order to first order,
changing the fitting range, and fixing the number of events for the
$J/\psi \to \gamma \eta_c$ and $\psi(3686) \to \gamma \chi_{c0},
\chi_{c0} \to \gamma J/\psi$ to the numbers expected, and the result
for each case is compared with our standard fit to determine the
systematic uncertainty for that case.

For the alternative strategy, the $\psi(3686) \to \gamma \chi_{c1}$ and
$\gamma \chi_{c2}$ peaks are described by asymmetric Gaussians with
Crystal Ball tails on both sides.  The other signal peaks and
backgrounds are the same.  A simultaneous fit is done to better
constrain the asymmetric Gaussian and Crystal Ball tail parameters,
which are common between the inclusive and exclusive distributions.

For the $\psi(3686) \to \gamma \chi_{cJ}$ systematic uncertainties,
the fitting range is changed from 0.08 -- 0.5 GeV to 0.08 -- 0.35 GeV,
which removes the $\chi_{cJ} \to \gamma J/\psi$ peaks and changes the
number of parameters used in the fit.  For the $\chi_{cJ} \to \gamma
J/\psi$ systematic uncertainties, the range is changed from 0.08 -- 0.5
GeV to 0.2 -- 0.54 GeV, which removes the $\psi(3686) \to \gamma
\chi_{c1}$ and $\psi(3686) \to \gamma \chi_{c2}$ peaks and produces a
rather large systematic uncertainty due to the background in the fit
of data preferring a pure polynomial background in the latter case.
The total systematic uncertainties from fitting for each branching
fraction are determined by adding the systematic uncertainties from
each source in quadrature.


The signal for $\psi(3686) \to \gamma \chi_{c0}, \chi_{c0} \to \gamma
J/\psi$ is very small and sits on the tail of $\psi(3686) \to \gamma
\chi_{c0}$.  It is therefore difficult to fit this peak as indicated
by the very large fitting systematic uncertainty for this process.

\subsection{MC Signal Shape}

The signal shapes used in fitting the photon energy distribution are
determined by matching MC radiative photons with reconstructed photons
in the EMC, where the angle between the photons is required to be less
than $\Delta \theta = 0.08$ radians.  This selection could bias the
signal shapes used in the fitting.  The systematic uncertainty
associated with this selection is determined by changing the $\Delta
\theta$ selection requirement to 0.04 radians.  The differences for
each decay are taken as the systematic uncertainties in the signal
shape.

\subsection{\boldmath Higher order multipoles for $\psi(3686) \to \gamma
  \chi_{c1}$ and $\chi_{c2}$}
\label{higher}

Angular distributions for $\psi(3686) \to \gamma \chi_{cJ}$ are
generated according to those expected for $E1$ radiative transitions.
This approach is accurate enough for $\psi(3686) \to \gamma
\chi_{c0}$, but higher order multipole contributions must be
considered for $\psi(3686) \to \gamma \chi_{c1}$ and $\psi(3686) \to
\gamma \chi_{c2}$ decays. Also the angular distributions for
$\chi_{cJ} \to \gamma J/\psi$ MC events do not agree with data.
BESIII has measured the angular distributions for $\psi(3686) \to
\gamma \chi_{cJ}, \chi_{cJ} \to \gamma J/\psi$~\cite{bam158}, and
these distributions have been fitted to $1 + \alpha \cos^2 \theta$,
where $\theta$ is the laboratory polar angle, and the values of
$\alpha$ have been determined.  Using these values of $\alpha$, it is
possible to calculate the differences in the geometric acceptance
between data and the inclusive $\psi(3686)$ MC.  The acceptance
efficiency for a given value of $\alpha$ is given by the integral of
$1 + \alpha\cos^2 \theta$ from $\cos \theta = -0.8$ to $\cos \theta =
0.8$ divided by the integral between $-1$ and $+1$.
Using the values of $\alpha$ that were used to generate the MC events
and those obtained based on Ref.~\cite{bam158}, the changes in the
efficiencies are 0.61\,\% for $\psi(3686) \to \gamma \chi_{c1}$ and
0.60\,\% for $\psi(3686) \to \gamma \chi_{c2}$.  For $\chi_{cJ} \to
\gamma J/\psi$ $(J = 1,2)$, the changes are 0.35\,\% and 3.82\,\%,
respectively.  The changes to the branching fractions from the changes
in efficiencies are taken as the systematic uncertainties due to the
higher order multipole corrections.


\subsection{\boldmath $|\cos \theta| < 0.8$}
The systematic uncertainty associated with the $|\cos \theta| < 0.8$
requirement is determined by using the requirement $|\cos \theta| <
0.75$ instead and by comparing the results with the standard
requirement.  This tests whether there are edge effects with the EMC
that are not fully modeled by the MC simulations.

\subsection{Event weighting}
As described in Section~\ref{compare}, MC events are weighted to give
better agreement for the $\pi^0$ distributions between data and MC
simulation, as well as to include the $E1$ transition $E_{\gamma}^3$
weight.  The systematic uncertainty associated with the $w_{\pi^0}$
weight is determined by turning off its weighting and taking the
difference in results as the systematic uncertainties.

\subsection{Continuum energy difference}
\label{conenedif}
Data distributions, including the inclusive photon energy distribution
for data, are defined as data minus scaled continuum data.  While this
takes into consideration the effect on the normalization of the
continuum due to the difference in luminosity and energy, it does not
consider the difference in the energy scale of the photons.  To
determine the systematic uncertainty due to this effect, the photon
energies of the continuum data were scaled by the ratio of the CM
energies, 3.686/3.65, and the scaled distribution was subtracted from
data, and the fitting redone. The differences with respect to the
standard analysis are taken as the systematic uncertainties of this
effect.

\subsection{Other systematic uncertainties}

The photon detection efficiency is studied utilizing the control
samples $\psi(3686) \to \pi^+\pi^- J/\psi, J/\psi \to \rho^0\pi^0$ and
$\psi(3686) \to \pi^0\pi^0 J/\psi$ with $J/\psi \to l^+l^-$ and
$\rho^0\pi^0$.  The corresponding systematic uncertainty is estimated
by the difference of detection efficiency between data and MC samples,
and 1\,\% is assigned for each photon~\cite{photon_eff}.

The trigger efficiency is assumed to be very close to 100\,\% with
negligible uncertainty, since the average charged particle and photon
multiplicities are high. The number of $\psi(3686)$ events is $(106.41
\pm 0.86) \times 10^6$, which is obtained by studying inclusive
$\psi(3686)$ decays~\cite{Npsip}.  The uncertainties from all above
sources and the total systematic uncertainty, obtained by adding all
uncertainties quadratically, are listed in Table~\ref{systematics}.
Since the fitting uncertainty for $\psi(3686) \to \gamma \chi_{c0},
\chi_{c0} \to \gamma J/\psi$ is so large, indicating that this fit is
not very meaningful, only this uncertainty is listed in the table.

\section{Results}
\label{results}

Our results are listed in Table~\ref{final results}.  We also
calculate ratios of branching fractions, where common systematic
uncertainties cancel
\begin{eqnarray}
 \mathcal{B}(\psi(3686) \to \gamma \chi_{c0})/\mathcal{B}(\psi(3686) \to \gamma
\chi_{c1}) \nonumber \\ 
= 0.948 \pm 0.002 \pm 0.044 \nonumber \\
 \mathcal{B}(\psi(3686) \to \gamma \chi_{c0})/\mathcal{B}(\psi(3686) \to \gamma
\chi_{c2}) \nonumber \\
= 0.976 \pm 0.002 \pm 0.040 \nonumber \\ 
 \mathcal{B}(\psi(3686) \to \gamma \chi_{c2})/\mathcal{B}(\psi(3686) \to \gamma
\chi_{c1})  \nonumber \\
= 0.971 \pm 0.002 \pm 0.040 \nonumber
\end{eqnarray}

\begin{table*}[htb!]
\centering
\caption{Our branching fraction results, other results, and PDG
  compilation results.}
\vspace{0.05 in}
\begin{tabular}{l|c|c|c|c} \hline \hline
\T Branching Fraction & This analysis (\%) & Other (\%) & PDG~\cite{PDG16}
(\%)& PDG~\cite{PDG16} (\%)\\ 
 & & & Average & Fit \B \\ \hline
\T $\mathcal{B}(\psi(3686) \to \gamma \chi_{c0})$ & $9.389 \pm 0.014 \pm 0.332$ &
$9.22\pm0.11\pm0.46$~\cite{cleo04} & $9.2\pm0.4$ &$9.99 \pm 0.27$\\  
$\mathcal{B}(\psi(3686) \to \gamma \chi_{c1})$ & $9.905 \pm 0.011 \pm 0.353$ &
$9.07 \pm 0.11\pm 0.54$~\cite{cleo04} & $ 8.9 \pm 0.5$ & $9.55 \pm 0.31$\\  
$\mathcal{B}(\psi(3686) \to \gamma \chi_{c2})$ &$ 9.621 \pm 0.013\pm0.272 $ &
$9.33\pm0.14\pm0.61$~\cite{cleo04} & $8.8\pm0.5$ &$9.11 \pm 0.31$ \B \\ \hline 
\T $\mathcal{B}(\psi(3686) \to \gamma \chi_{c0})\times
\mathcal{B}(\chi_{c0} \to \gamma J/\psi)$ & $ 0.024 \pm 0.015 \pm
0.205$& $0.125 \pm 0.007 \pm 0.013$~\cite{mendez}& $ 0.131\pm0.035$ &
$0.127\pm0.006$ \\  
 & & $0.151 \pm 0.003 \pm 0.010$~\cite{xiaorui}& & \\  
   & & $0.158\pm0.003\pm0.006$~\cite{bam158} & & \\
$\mathcal{B}(\psi(3686) \to \gamma \chi_{c1})\times
\mathcal{B}(\chi_{c1} \to \gamma J/\psi)$ & $ 3.442 \pm 0.010 \pm
0.132$& $3.56 \pm 0.03 \pm 0.12$~\cite{mendez}& $2.93\pm0.15$ & $ 3.24\pm0.07$\\  
 & & $3.377 \pm 0.009 \pm 0.183$~\cite{xiaorui}& & \\  
   & & $3.518\pm0.01\pm0.120$~\cite{bam158} & & \\
$\mathcal{B}(\psi(3686) \to \gamma \chi_{c2})\times
\mathcal{B}(\chi_{c2} \to \gamma J/\psi)$ & $ 1.793 \pm 0.008 \pm
0.163$ & $1.95\pm0.02\pm0.07$~\cite{mendez}& $1.52\pm0.15$ & $1.75 \pm 0.04$ \\ 
& & $1.874\pm0.007\pm0.102$~\cite{xiaorui}& &  \\ 
   & & $1.996\pm0.008\pm0.070$~\cite{bam158} & & \B \\ \hline
\T $\mathcal{B}(\chi_{c0} \to \gamma J/\psi)$ & $0.25 \pm 0.16 \pm 2.15$ & $2 \pm 0.2 \pm 0.2$~\cite{adam05a} && $1.27
\pm 0.06 $\\  
$\mathcal{B}(\chi_{c1} \to \gamma J/\psi)$ & $34.75 \pm 0.11 \pm 1.70$ & $37.9 \pm 0.8 \pm 2.1$~\cite{adam05a} && $33.9
\pm 1.2 $\\  
$\mathcal{B}(\chi_{c2} \to \gamma J/\psi)$ & $18.64 \pm 0.08 \pm 1.69$ & $19.9 \pm 0.5 \pm 1.2$~\cite{adam05a} &&
$19.2 \pm 0.7 \B $\\  
  \hline \hline
\end{tabular}
\label{final results}
\end{table*}

For comparison with some theoretical calculations, we also determine
partial widths using our branching fractions and the world average
full widths~\cite{PDG16}.  Table~\ref{widths} contains our partial
width results, as well as theoretical predictions, reproduced from
Table VI in Ref.~\cite{Deng1}. The theoretical predictions include the
linear potential (LP) and screened potential (SP) models~\cite{Deng1},
as well as earlier predictions from a relativistic quark model
(RQM)~\cite{RQM}, non-relativistic potential and Godfrey-Isgur
relativized potential models (NR/GI)~\cite{NR/GI}, and color screened
models, calculated with zeroth order wave functions (SNR$_0$) and
first order relativistically corrected wave functions
(SNR$_1$)~\cite{SNR}.

\begin{table*}[tb]
\centering
\caption{Partial widths (keV) of radiative transitions for $\psi(3686)
  \to \gamma J/\psi$ and $\chi_{cJ} \to \gamma J/\psi$.  Shown are
  our experimental results and predictions from a relativistic quark
  model (RQM)~\cite{RQM}; non-relativistic potential and Godfrey-Isgur relativized potential
  models (NR/GI)~\cite{NR/GI}; color screened  models~\cite{SNR}, calculated with  zeroth order wave functions (SNR$_0$) and
  first order relativistically  corrected wave functions 
  (SNR$_{1}$); and linear potential
  (LP) and screened potential models (SP)~\cite{Deng1}.  The $\Gamma_{E1}$ predictions include only $E1$ transition calculations, while the $\Gamma_{EM}$ results include higher order multipole corrections.}
\vspace{0.05 in}
\begin{tabular}{c|c|ccccc|ccc} \hline \hline
\T Initial & Final & \multicolumn{5}{c|}{$\Gamma_{E1}$ (keV)} & \multicolumn{3}{c}{$\Gamma_{EM}$ (keV)} \B  \\ \cline{3-10} 
\T state   & state & RQM~\cite{RQM} & NR/GI~\cite{NR/GI} & SNR$_{0/1}$~\cite{SNR} & LP~\cite{Deng1} & SP~\cite{Deng1} & LP~\cite{Deng1} & SP~\cite{Deng1} & This analysis \B \\ \hline 
\T $\psi(3686)$ & $\chi_{c0}$ & 26.3 & 63/26 & 74/25 & 27 & 26 & 22 & 22 & $26.9 \pm 1.8$ \\
           & $\chi_{c1}$ & 22.9 & 54/29 & 62/36 & 45 & 48 & 42 & 45 & $28.3\pm1.9$ \\
           & $\chi_{c2}$ & 18.2 & 38/24 & 43/34 & 36 & 44 & 38 & 46 & $ 27.5\pm1.7$ \B \\ \hline
\T $\chi_{c0}$ & $J/\psi$   & 121 & 152/114& 167/117 & 141 & 146 & 172& 179  &\\
$\chi_{c1}$ &            & 265 & 314/239& 354/244 & 269 & 278 & 306 & 319 & $306 \pm 23$ \\
$\chi_{c2}$ &            & 327 & 424/313 & 473/309 & 327 & 338& 284 & 292 & $ 363\pm41$ \B \\ \hline  \hline
\end{tabular}
\label{widths}
\end{table*}

\section{Summary}

Our results, CLEO measurements~\cite{cleo04,mendez,adam05a}, previous
BESIII measurements~\cite{xiaorui,bam158}, and PDG
results~\cite{PDG16} are listed in Table~\ref{final results}.  Our
$\psi(3686) \to \gamma \chi_{cJ}$ branching fractions are the most
precise.  The branching fractions for $\psi(3686) \to \gamma
\chi_{cJ}$ agree with CLEO within one standard deviation, except for
$\psi(3686) \to \gamma \chi_{c1}$ which differs by 1.3 standard
deviations.  The product branching fractions $\mathcal{B}(\psi(3686)
\to \gamma \chi_{c1})\times \mathcal{B}(\chi_{c1} \to \gamma J/\psi)$
and $\mathcal{B}(\psi(3686) \to \gamma \chi_{c2})\times
\mathcal{B}(\chi_{c2} \to \gamma J/\psi)$ agree with the previous
BESIII measurements. Because of the difficulty in fitting $\psi(3686)
\to \gamma \chi_{c0}, \chi_{c0} \to \gamma J/\psi$, our product
branching fraction has a very large systematic error compared with
those using exclusive decays.

Partial widths are shown in Table~\ref{widths}.  For comparison with
models, experimental results have become accurate enough (partly due
to this measurement) to become sensitive to fine details of the
potentials, e.g. relativistic effects, screening effects, and higher
partial waves.


\clearpage
\section{Acknowledgments}

The BESIII collaboration thanks the staff of BEPCII and the IHEP
computing center for their strong support. This work is supported in
part by National Key Basic Research Program of China under Contract
No. 2015CB856700; National Natural Science Foundation of China (NSFC)
under Contracts Nos. 11235011, 11322544, 11335008, 11425524, 11635010;
the Chinese Academy of Sciences (CAS) Large-Scale Scientific Facility
Program; the CAS Center for Excellence in Particle Physics (CCEPP);
the Collaborative Innovation Center for Particles and Interactions
(CICPI); Joint Large-Scale Scientific Facility Funds of the NSFC and
CAS under Contracts Nos. U1232201, U1332201; CAS under Contracts
Nos. KJCX2-YW-N29, KJCX2-YW-N45; 100 Talents Program of CAS; National
1000 Talents Program of China; INPAC and Shanghai Key Laboratory for
Particle Physics and Cosmology; German Research Foundation DFG under
Contracts Nos. Collaborative Research Center CRC 1044, FOR 2359;
Istituto Nazionale di Fisica Nucleare, Italy; Joint Large-Scale
Scientific Facility Funds of the NSFC and CAS under Contract
No. U1532257; Joint Large-Scale Scientific Facility Funds of the NSFC
and CAS under Contract No. U1532258; Koninklijke Nederlandse Akademie
van Wetenschappen (KNAW) under Contract No. 530-4CDP03; Ministry of
Development of Turkey under Contract No. DPT2006K-120470; National
Natural Science Foundation of China (NSFC) under Contract
No. 11575133; National Science and Technology fund; NSFC under
Contract No. 11275266; The Swedish Resarch Council; U. S. Department
of Energy under Contracts Nos. DE-FG02-05ER41374, DE-SC-0010504,
DE-SC0012069; University of Groningen (RuG) and the Helmholtzzentrum
fuer Schwerionenforschung GmbH (GSI), Darmstadt; WCU Program of
National Research Foundation of Korea under Contract
No. R32-2008-000-10155-0

\end{document}